\documentclass[a4paper,11pt]{article}
\pdfoutput=1 

\usepackage{jheppub} 

\usepackage[T1]{fontenc} 
\usepackage{amssymb, bm, graphicx, graphics, color, mathrsfs, multirow, amsmath}
\usepackage{subfigure} \subfiguretopcaptrue

\newcommand{\ben}{\begin{eqnarray}}
\newcommand{\een}{\end{eqnarray}}

\newcommand{\bp}{\bar\phi}
\newcommand{\Pm}{_{;\mu}}
\newcommand{\Pn}{_{;\nu}}
\newcommand{\Pu}{_{,u}}
\newcommand{\Pv}{_{,v}}
\newcommand{\Puv}{_{,uv}}

\newcommand{\bz}{\bar{z}}
\newcommand{\bs}{\bar{s}}
\newcommand{\bw}{\bar{w}}

\makeatletter
\newcommand{\raisemath}[1]{\mathpalette{\raisem@th{#1}}}
\newcommand{\raisem@th}[3]{\raisebox{#1}{$#2#3$}}
\makeatother

\title{\boldmath Scalar field as~an~intrinsic time measure in~coupled dynamical matter-geometry systems.\\ II. Electrically charged gravitational collapse}

\author[a,b]{Anna Nakonieczna}
\author[c]{and Dong-han Yeom}

\affiliation[a]{Institute of~Physics, Maria Curie-Sk{\l}odowska University, \\
Plac Marii Curie-Sk{\l}odowskiej 1, 20-031 Lublin, Poland}
\affiliation[b]{Institute of~Agrophysics, Polish Academy of~Sciences,\\
Do{\'s}wiadczalna 4, 20-290 Lublin, Poland}
\affiliation[c]{Leung Center for~Cosmology and~Particle Astrophysics, National Taiwan University,\\
No. 1, Sec. 4, Roosevelt Road, Taipei 10617, Taiwan}

\emailAdd{aborkow@kft.umcs.lublin.pl}
\emailAdd{innocent.yeom@gmail.com}

\abstract{
Investigating the~dynamics of~gravitational systems, especially in~the~regime of~quantum gravity, poses a~problem of~measuring time during the~evolution. One~of~the~approaches to this issue is using one of~the~internal degrees of~freedom as~a~time variable. The~objective of~our~research was to~check whether a~scalar field or any other dynamical quantity being a~part of~a~coupled multi-component matter-geometry system can be treated as~a~`clock' during its evolution. We~investigated a~collapse of~a~self-gravitating electrically charged scalar field in~the~Einstein and~Brans-Dicke theories using the~2+2 formalism. Our~findings concentrated on~the~spacetime region of~high curvature existing in~the~vicinity of~the~emerging singularity, which is essential for the~quantum gravity applications. We~investigated several values of~the~Brans-Dicke coupling constant and~the~coupling between the~Brans-Dicke and~the~electrically charged scalar fields. It~turned out that both~evolving scalar fields and~a~function which measures the~amount of~electric charge within a~sphere of~a~given radius can be used to quantify time nearby the~singularity in~the~dynamical spacetime part, in~which the~apparent horizon surrounding the~singularity is spacelike. Using them in~this respect in~the~asymptotic spacetime region is possible only when both fields are present in~the~system and, moreover, they are coupled to each other. The~only nonzero component of~the~Maxwell field four-potential cannot be used to quantify time during the~considered process in~the~neighborhood of~the~whole central singularity. None of~the~investigated dynamical quantities is a~good candidate for measuring time nearby the~Cauchy horizon, which is also singular due to the~mass inflation phenomenon.
}

\begin{document} 
\maketitle
\flushbottom

\section{Introduction}
\label{sec:intro}

Time measuring in~dynamical gravitational systems is an~important and~demanding issue, especially when one considers investigating them within the~quantized canonical formulations of~the~theory of~gravity. Any general notion of~a~time measurer which could be transferred from the~classical to the~quantum level has not yet been proposed. One of~the~ideas in~this regard, which has been widely used in~analyses within the~fields of~canonical gravity and~cosmology, is to employ one of~the~internal degrees of~freedom of~a~time-dependent system to act as~a~`clock'~\cite{DeWitt1967}. However, arguments in~favor of~such a~treatment are limited to certain cases and~thus detailed investigations are still required. The~current research addresses the~problem of~time quantification with the~use of~scalar fields and~also other dynamical quantities present in~evolving coupled multi-component matter-geometry systems. The~studied evolution was a~gravitational collapse of~an~electrically charged scalar field in~the~Einstein and~Brans-Dicke theories.

The~gravitational collapse of~a~self-interacting electrically charged scalar field is a~toy-model of~a~more realistic collapse, which produces the~rotating and~neutral Kerr black hole. It~leads to the~formation of~a~dynamical Reissner-Nordstr{\"o}m spacetime, which possesses a~spacelike central singularity surrounded by the~null Cauchy and~event horizons~\cite{HodPiran1997-3485,HodPiran1998-1554,HodPiran1998-1555,OrenPiran2003-044013}. The~influence of~pair creation in~strong electric fields on~the~outcomes of~the~process was described in~\cite{SorkinPiran2001-084006,SorkinPiran2001-124024}. Its course when the~neutralization and~the~black hole evaporation due to the~Hawking radiation emission are taken into account was studied in~\cite{HongHwangStewartYeom2010-045014,HwangYeom2011-064020}. The~evolution of~interest was also examined in~the~dilaton~\cite{BorkowskaRogatkoModerski2011-084007,NakoniecznaRogatko2012-3175}, phantom~\cite{NakoniecznaRogatkoModerski2012-044043} and~Brans-Dicke~\cite{HansenYeom2014-040,HansenYeom2015-019} theories of~gravity. The~course and~results of~the~electrically charged gravitational collapse of~a~scalar field in~the~de~Sitter spacetime were characterized in~\cite{ZhangZhangZouWang2016-064036}.

The~current paper describes the~continuation of~the~studies whose outcomes were presented in~\cite{NakoniecznaYeom2016-049}, which from now on~will be referred to~as~paper~I. The~performed analyses dealt with the~problem of~a~dynamical gravitational collapse of~neutral coupled matter-geometry systems in~the~context of~performing time measurements intrinsically during the~process. A~broad discussion on~the~following topics can be found in~paper~I:
\begin{itemize}
 \item the~existing approaches to intrinsic time measurements in~dynamical gravitational systems and~their specific implementations in~quantum gravity and~cosmology,
 \item a~discussion on~the~arguments in~favor of~the~above propositions and~the~justification of~the~undertaken studies in~this context,
 \item a~synopsis to the~Brans-Dicke theory of~gravity, its relations to experimental data, the~Einstein theory of~relativity and~cosmology,
 \item a~justification for choosing the~Brans-Dicke setup for the~conducted research and~a~brief summary of~previous numerical achievements within the~theory,
 \item specific arguments for choosing the~particular values of~the~free evolution parameters (the Brans-Dicke coupling constant $\omega$ and~the~coupling between the~Brans-Dicke and~scalar fields $\beta$), which were used during the~analyses.
\end{itemize}
Thus, in~order to get acquainted with the~above-listed essential issues related to the~background and~the~core of~our research, we refer the~Reader to paper~I.

As~was pointed out at the~beginning of~the~introduction, the~issue of~using a~dynamical quantity present in~the~coupled matter-geometry system as~an~intrinsic `clock' during inspecting its evolution is crucial for the~spacetime regions of~high curvature. For~this reason, the~discussion on~the~results will mainly concentrate on~the~neighborhood of~spacetime singularities, which emerge during the~gravitational collapse of~matter. In~order to treat the~particular quantity as~a~time measurer, its constancy hypersurfaces must fulfill two conditions, at least within the~spacetime regions of~interest. First, the~slices have to be spacelike in~these areas. Second, their parametrization needs to remain monotonic during the~whole evolution.

The~current paper was constructed in~the~following way. Section~\ref{sec:model} contains the~description of~the~theoretical formulation of~the~investigated problem. The~necessary details of~numerical computations and~the~results presentation are placed in~section~\ref{sec:details}. The~first general aim of~our analyses was to~investigate the~potential of~measuring time with the~use of~dynamical quantities during the~collapse of~a~self-gravitating electrically charged scalar field within the~Einstein theory. The~related results are presented in~section~\ref{sec:ein}. The~second general goal was to address the~above problem in~the~context of~a~dynamical gravitational evolution in~the~Brans-Dicke theory. The~obtained outcomes are elaborated in~section~\ref{sec:bd}. The~overall conclusions are gathered in~section~\ref{sec:conclusions}, while appendix~\ref{sec:appendix} contains a~comment on~the~numerical computations and~the~code tests.

\section{Gravitational evolution of~an~electrically charged scalar field}
\label{sec:model}

\subsection{Covariant form of~the~equations of~motion}
\label{sec:model-cov}

The~action which describes an~electrically charged scalar field in~the~Brans-Dicke theory with a~nontrivial exponential coupling between the~two scalar fields present within the~system is
\ben\label{eqn:BDaction}
S_{BD}=\int d^4 x \sqrt{-g} \left[ \frac{1}{16\pi} 
\left( \Phi R - \frac{\omega}{\Phi} \Phi\Pm \Phi\Pn g^{\mu\nu} \right) + \Phi^\beta L^{EM} \right],
\een
where $g$ denotes the~determinant of~the~metric $g^{\mu\nu}$, $R$ is the~Ricci scalar, $\Phi$ and~$\omega$ are the~Brans-Dicke field function and~coupling constant, respectively, while $\beta$ is a~constant which controls the~coupling between the~Brans-Dicke and~electrically charged fields. The~Lagrangian of~the~latter field has the~usual form
\ben\label{eqn:BDaction-lag}
L^{EM}=-\frac{1}{2} \left( \phi\Pm + ieA_\mu\phi \right) \left( \bp\Pn - ieA_\nu\bp \right) g^{\mu\nu} 
- \frac{1}{16\pi} F_{\mu\nu}F^{\mu\nu},
\een
in which the~complex field $\phi$ is charged under a~$U(1)$ gauge field, whose four-potential is denoted as~$A_\mu$ and~the~coupling between the~two is $e$. The~quantity $F_{\mu\nu}\equiv A_{\nu;\mu}-A_{\mu;\nu}$ is the~strength tensor of~the~gauge field, while $i$ is the~imaginary unit.

The~equations of~motion of~the~gravitational field resulting from the~above theoretical setup can be written as~follows:
\ben\label{eqn:ein}
G_{\mu\nu}=8\pi \left( T_{\mu\nu}^{BD} + \Phi^{\beta-1} T_{\mu\nu}^{EM} \right) \equiv 8\pi T_{\mu\nu}.
\een
The~components of~the~Einstein tensor $G_{\mu\nu}$ are determined by~the~selected metric and~their form will be presented in~the~next section. The~stress-energy tensors of~the~Brans-Dicke and~electrically charged fields are
\ben\label{eqn:set-bd}
T_{\mu\nu}^{BD} &=& \frac{1}{8\pi\Phi} \left( \Phi_{;\mu\nu} - g_{\mu\nu}\Phi_{;\rho\sigma} g^{\rho\sigma} \right)
+ \frac{\omega}{8\pi\Phi^2} \left( \Phi\Pm \Phi\Pn - \frac{1}{2} g_{\mu\nu} \Phi_{;\rho} \Phi_{;\sigma} g^{\rho\sigma} \right),\\
\label{eqn:set-sf}
T_{\mu\nu}^{EM} &=& \frac{1}{2} \left( \phi\Pm \bp\Pn + \bp\Pm \phi\Pn \right)
+\frac{1}{2} \left( \bp\Pm ie A_\nu \phi + \bp\Pn ie A_\mu \phi - \phi\Pm ie A_\nu \bp - \phi\Pn ie A_\mu \bp \right) + \nonumber\\
&& + \frac{1}{4\pi} F_{\mu\rho} F_\nu^{\ \rho} + e^2 A_\mu A_\nu \phi\bp +g_{\mu\nu} L^{EM}.
\een

The~covariant forms of~the~equations of~motion of~the~Brans-Dicke field, the~electrically charged scalar field and~its complex conjugate and~the~Maxwell field are the~following:
\ben\label{eqn:BD}
\Phi_{;\mu\nu} g^{\mu\nu} - \frac{8\pi\Phi^\beta}{3+2\omega} \left( T^{EM} - 2\beta L^{EM} \right) &=& 0, \\
\label{eqn:sf1}
\phi_{;\mu\nu} g^{\mu\nu} + ieA^\mu \left( 2\phi\Pm + ieA_\mu\phi \right) + ie A_{\mu;\nu} g^{\mu\nu} \phi
+ \frac{\beta}{\Phi} \Phi\Pm \left( \phi\Pn + ieA_\nu\phi \right) g^{\mu\nu} &=& 0, \\
\label{eqn:sf2}
\bp_{;\mu\nu} g^{\mu\nu} - ieA^\mu \left( 2\bp\Pm - ieA_\mu\bp \right) - ie A_{\mu;\nu} g^{\mu\nu} \bp
+ \frac{\beta}{\Phi} \Phi\Pm \left( \bp\Pn - ieA_\nu\bp \right) g^{\mu\nu} &=& 0, \\
\label{eqn:em}
\frac{1}{2\pi} \left( F_{\ \mu;\nu}^\nu + \frac{\beta}{\Phi} F_{\ \mu}^\nu \Phi\Pn \right)
- ie\phi \left( \bp\Pm - ieA_\mu\bp \right) + ie\bp \left( \phi\Pm + ieA_\mu\phi \right) &=& 0,
\een
where $T^{EM}$ denotes the~trace of~\eqref{eqn:set-sf}.

\subsection{Dynamics in~double null coordinates}
\label{sec:model-dn}

The~evolution will be traced in~double null coordinates ($u$,~$v$, $\theta$, $\varphi$), in~which a~spherically symmetric line element is
\ben\label{eqn:metric}
ds^2=-\alpha\left(u,v\right)^2 dudv +r\left(u,v\right)^2 d\Omega^2,
\een
where~$u$ and~$v$ are retarded and~advanced times, respectively, $d\Omega^2=d\theta^2+\sin^2\theta d\varphi^2$ is the~line element of~a~unit sphere, while $\theta$ and~$\varphi$ denote angular coordinates. The~arbitrary functions $\alpha$ and~$r$ depend on~both the~retarded and~advanced time. Their dynamics reflects the~evolution of~spacetime in~the~investigated system.

The~rescaling of~the~complex field function $s\equiv\sqrt{4\pi}\phi$ simplifies the~form of~the~equations of~motion. The~second-order differential equations were turned into the~first-order ones via the~substitutions
\ben\label{eqn:subst}
\begin{split}
h &=\frac{\alpha\Pu}{\alpha}, \quad d=\frac{\alpha\Pv}{\alpha}, \quad f=r\Pu, \quad g=r\Pv,\\
W &=\Phi\Pu, \quad Z=\Phi\Pv, \quad w=s\Pu, \quad z=s\Pv.
\end{split}
\een

The~components of~the~Einstein tensor related to~the~line element~\eqref{eqn:metric} are
\ben
G_{uu} &=& -\frac{2}{r} \left( f\Pu - 2fh \right), \\
G_{vv} &=& -\frac{2}{r} \left( g\Pv - 2gd \right), \\
G_{uv} &=& \frac{1}{2r^2} \left( 4rf\Pv + \alpha^2 + 4fg \right), \\
G_{\theta\theta} &=& \sin^{-2}\theta\ G_{\varphi\varphi}=-\frac{4r^2}{\alpha^2} \left( d\Pu + \frac{f\Pv}{r} \right),
\een
while the~elements of~the~stress-energy tensors of~the~considered theory~\eqref{eqn:set-bd} and~\eqref{eqn:set-sf} are the~following:
\ben
T_{uu}^{BD} &=& \frac{1}{8\pi\Phi} \left( W\Pu - 2hW \right) + \frac{\omega}{8\pi\Phi^2}W^2, \\
T_{vv}^{BD} &=& \frac{1}{8\pi\Phi} \left( Z\Pv - 2dZ \right) + \frac{\omega}{8\pi\Phi^2}Z^2, \\
T_{uv}^{BD} &=& -\frac{Z\Pu}{8\pi\Phi} - \frac{gW + fZ}{4\pi r\Phi}, \\
T_{\theta\theta}^{BD} &=& \sin^{-2}\theta\ T_{\varphi\varphi}^{BD}=\frac{r^2}{2\pi\alpha^2 \Phi}Z\Pu
+ \frac{r}{4\pi\alpha^2 \Phi} \left( gW + fZ \right) + \frac{\omega r^2}{4\pi\Phi^2\alpha^2}WZ, \\
T_{uu}^{EM} &=& \frac{1}{4\pi} \left[ w\bw + ieA_u \left( \bw s - w\bs \right) + e^2A_u^2s\bs \right], \\
T_{vv}^{EM} &=& \frac{1}{4\pi} z\bz, \\
T_{uv}^{EM} &=& \frac{A_{u,v}^2}{4\pi\alpha^2}, \\
T_{\theta\theta}^{EM} &=& \sin^{-2}\theta\ T_{\varphi\varphi}^{EM}=\frac{r^2}{4\pi\alpha^2}
\left[ w\bz + z\bw + ieA_u \left( \bz s - z\bs \right) + \frac{2A_{u,v}^2}{\alpha^2} \right].
\een
Due to~the~gauge freedom fixing, the~only nonzero component of~the~electromagnetic four-potential is $A_u$~\cite{OrenPiran2003-044013}, which is a~function of~advanced and~retarded times.

The~$\theta\theta$ (or $\varphi\varphi$) and~$uv$ components of~the~Einstein equations~\eqref{eqn:ein} can be written together with the~equation of~the~Brans-Dicke field~\eqref{eqn:BD} in~a~matrix form
\ben\label{eqn:matrix1}
\begin{pmatrix}
1\ & \frac{1}{r}\ & \frac{1}{\Phi} \\
0\ & 1\ & \frac{r}{2\Phi} \\
0\ & 0\ & r
\end{pmatrix}
\begin{pmatrix}
d\Pu \\
f\Pv \\
Z\Pu
\end{pmatrix}
=
\begin{pmatrix}
\mathcal{A} \\
\mathcal{B} \\
\mathcal{C}
\end{pmatrix}.
\een
The~elements of~the~right-hand side vector are
\ben
\mathcal{A} &\equiv& -\frac{2\pi\alpha^2}{r^2\Phi} \widetilde{T}_{\theta\theta}^{EM} - \frac{1}{2r\Phi} \left( gW+fZ \right)
- \frac{\omega}{2\Phi^2}WZ, \\
\mathcal{B} &\equiv& -\frac{\alpha^2}{4r} - \frac{fg}{r} + \frac{4\pi r}{\Phi} \widetilde{T}_{uv}^{EM}
- \frac{1}{\Phi} \left( gW+fZ \right), \\
\mathcal{C} &\equiv& -fZ - gW - \frac{2\pi r\alpha^2}{3+2\omega} \left( \widetilde{T}^{EM} - 2\beta\widetilde{L}^{EM} \right),
\een
where $\widetilde{Q}\equiv\Phi^\beta Q$ for any quantity $Q$ and
\ben
T^{EM} &=& -\frac{4}{\alpha^2} T_{uv}^{EM} + \frac{2}{r^2} T_{\theta\theta}^{EM}, \\
L^{EM} &=& \frac{1}{4\pi\alpha^2} \left( w\bz + z\bw \right) + \frac{ieA_u}{4\pi\alpha^2} \left( \bz s - z\bs \right)
+ \frac{A_{u,v}^2}{2\pi\alpha^4}.
\een
An~equivalent form of~\eqref{eqn:matrix1}, appropriate for numerical computations, is
\ben\label{eqn:matrix2}
\begin{pmatrix}
d\Pu=h\Pv \\
g\Pu=f\Pv \\
Z\Pu=W\Pv
\end{pmatrix}
= \frac{1}{r^2}
\begin{pmatrix}
r^2\ & -r\ & -\frac{r}{2\Phi} \\
0\ & r^2\ & -\frac{r^2}{2\Phi} \\
0\ & 0\ & r
\end{pmatrix}
\begin{pmatrix}
\mathcal{A} \\
\mathcal{B} \\
\mathcal{C}
\end{pmatrix}.
\een

The~remaining of~the~Einstein equations~\eqref{eqn:ein}, i.e.,~their $uu$ and~$vv$ components, provide the~constraints
\ben\label{constr1}
f\Pu &=& 2fh - \frac{r}{2\Phi} \left( W\Pu - 2hW \right) - \frac{r\omega}{2\Phi^2}W^2
- \frac{4\pi r}{\Phi}\widetilde{T}_{uu}^{EM}, \\
\label{constr2}
g\Pv &=& 2dg - \frac{r}{2\Phi} \left( Z\Pv - 2dZ \right) - \frac{r\omega}{2\Phi^2}Z^2
- \frac{4\pi r}{\Phi}\widetilde{T}_{vv}^{EM}.
\een

The~evolution of~the~complex scalar field described by~\eqref{eqn:sf1} and~\eqref{eqn:sf2} is governed by~the~equations
\ben\label{eqn:sf1-dn}
z\Pu=w\Pv &=& -\frac{fz}{r} - \frac{gw}{r} - ieA_uz - \frac{ieA_ugs}{r} - \frac{ie}{4r^2}\alpha^2 qs - \frac{\beta}{2\Phi} \left( Wz + Zw + iesA_uZ \right),\qquad \\
\label{eqn:sf2-dn}
\bz\Pu=\bw\Pv &=& -\frac{f\bz}{r} - \frac{g\bw}{r} + ieA_u\bz + \frac{ieA_ug\bs}{r} + \frac{ie}{4r^2}\alpha^2 q\bs - \frac{\beta}{2\Phi} \left( W\bz + Z\bw - ie\bs A_uZ \right),
\een
where the~function reflecting the~amount of~electric charge contained within a~sphere of~a~radius $r\left(u,v\right)$ was defined as~\cite{OrenPiran2003-044013}
\ben\label{eqn:q}
q\left(u,v\right)=\frac{2r^2A_{u,v}}{\alpha^2}.
\een
The~Maxwell field dynamics~\eqref{eqn:em} is described by~the~relations
\ben\label{eqn:Au-dn}
A_{u,v} &=& \frac{\alpha^2 q}{2r^2}, \\
\label{eqn:q-dn}
q\Pv &=& -\frac{ier^2}{2} \left( \bs z-s\bz \right) - \beta q \frac{Z}{\Phi}.
\een

\section{Details of~computer simulations and~results analysis}
\label{sec:details}

\begin{figure}[tbp]
\centering
\subfigure[][]{\includegraphics[width=0.315\textwidth]{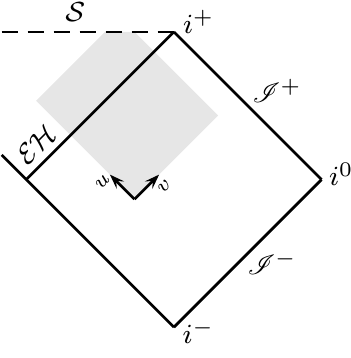}\label{fig:cd-s}}\hspace{2cm}
\subfigure[][]{\includegraphics[width=0.3\textwidth]{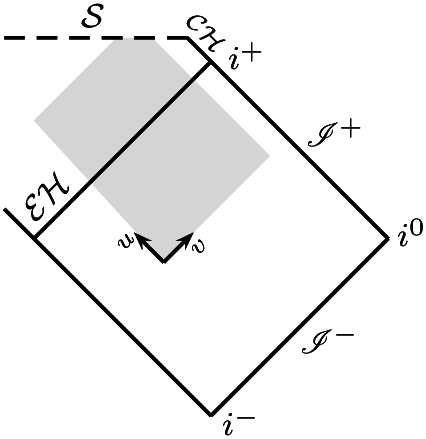}\label{fig:cd-rn}}
\caption{The~computational domain (marked gray) on~the~background of~the~Carter-Penrose diagram of~the~dynamical (a)~Schwarzschild and~(b)~Reissner-Nordstr\"{o}m spacetimes. The~central singularity along $r=0$, the~event and~Cauchy horizons are denoted as~$\mathcal{S}$, $\mathcal{EH}$ and~$\mathcal{CH}$, respectively. $\mathscr{I}^\pm$ and~$i^\pm$ are null and~timelike infinities, while $i^0$ is~a~spacelike infinity.}
\label{fig:domain}
\end{figure}

The~dynamics of~the~system containing an~electrically charged scalar field described by~\eqref{eqn:BDaction-lag} coupled with the~Brans-Dicke field according to~\eqref{eqn:BDaction} is governed in~double null coordinates by~a~set of~equations~\eqref{eqn:matrix2}, \eqref{eqn:sf1-dn}--\eqref{eqn:q-dn}. The~constraints, which were used to~control the~accuracy of~numerical calculations, are provided by~\eqref{constr1} and~\eqref{constr2}. The~computational domain, within which the~evolution was traced, is shown in~figure~\ref{fig:domain}. It~is presented against the~dynamical Schwarzschild and~Reissner-Nordstr\"{o}m spacetimes in~the~$(vu)$-plane. The~respective diagrams are relevant to~the~cases, in~which the~Cauchy horizon does not and~does form in~the~emerging spacetime. The~only freely specifiable initial conditions of~the~studied process were the~profiles of~the~complex scalar and~Brans-Dicke fields posed on~an~arbitrarily chosen $u=const.$ hypersurface, which were the~same as~in~paper~I. The~field functions were nonzero only within the~interval $v\in [0,20]$ and~thus the~spacetime region from within the~specified range will be referred to~as~dynamical. The~details of~the~numerical code and~its consistency tests are presented in~appendix~\ref{sec:appendix}.

The~investigated values of~the~evolution parameters ($\beta$ and~$\omega$) and~the~justification for their choice were discussed in~paper~I. $\delta\in[0,0.5]$ controls the~character of~the~scalar field, which is either real or~complex when the~parameter is equal and~not equal to~zero, respectively. Its~value was constant in~all computations of~the~current paper and~equal to~$0.5$, because obtaining a~charged field requires a~complex scalar coupled to~a~$U(1)$ gauge field. The~electric coupling constant was arbitrarily set as~$e=0.3$ in~all conducted simulations, as~changing its value (provided that it~was not equal to~zero) did not influence the~outcomes qualitatively, and~thus also the~ultimate conclusions. Spacetimes obtained in~the~case of~an~uncoupled collapse, i.e.,~for $e=0$, will be also presented for a~convenient comparison.

The~presentation of~the~results will be based on~the~Penrose spacetime diagrams ($r=const.$ lines in~the~$(vu)$-plane) and~plots of~constancy hypersurfaces of~adequate dynamical quantities ($|\phi|$, $\textrm{Re}\;\phi$, $\textrm{Im}\;\phi$, $A_u$ and~$q$), also within the~$(vu)$-plane. The~singularities and~future infinities are marked as~thick black curves on~the~plots and~signed, while the~apparent horizons, situated along the~hypersurfaces $r\Pv=0$ and~$r\Pu=0$, are marked as~red and~blue lines, respectively. The~areas, in~which the~constancy hypersurfaces of~the~particular quantity are spacelike are blue, and~the~regions with their timelike character are purple. In~most cases, the~constancy hypersurfaces were plotted only for the~real part of~the~complex scalar field. For~explanations, we refer the~Reader to~paper~I. In~order to~confirm the~fact that real and~imaginary parts of~the~complex field behave analogously and~to~check additional quantities, which may be of~interest during time measuring, the~constancy hypersurfaces of~the~field modulus, the~$u$-component of~the~Maxwell field four-potential and~the~charge function were plotted for selected cases. The~modulus of~the~field function was calculated according to~the~definition from the~values of~its real and~imaginary parts.

The~values of~the~evolution parameters used to~generate the~particular outcome will be presented above each diagram. The~ranges of~the~plotted quantities and~the~steps between adjacent lines representing their constant values are the~following.
\begin{itemize}
 \item On~the~diagrams of~spacetime structures $r$ differs from $r=0$ to~$r=40$, and~the~range is divided into $40$ steps.

 \item On~the~plots presenting the~constancy hypersurfaces of~the~real and~imaginary parts of~the~electrically charged scalar field, as~well as~its modulus, the~ranges of~$\textrm{Re}\;\phi$, $\textrm{Im}\;\phi$ and~$|\phi|$ from $-1.75$ to~$0.5$ are divided into $75$ steps.

 \item On~the~plots presenting the~constancy hypersurfaces of~the~Brans-Dicke field, its ranges, divided into $100$ steps in~each case, depend on~the~values of~$\omega$ and~$\beta$ as~follows.
 \vspace{-0.35cm}
 \begin{itemize}\renewcommand{\labelitemii}{$\cdot$}
  \item For~$\omega > -1.5$ and~$\beta=0$, the~range is from $\Phi=0$ to~$\Phi=1$.
  \item For~$\omega > -1.5$ and~$\beta=0.5$ and~$1$, $\Phi$ ranges from $0$ to~$10$.
  \item For~$\omega < -1.5$ and~$\beta=0$, $\Phi$ ranges from $1$ to~$20.9$.
  \item For~$\omega < -1.5$ and~$\beta=0.5$ and~$1$, the~range is from $\Phi=0$ to~$\Phi=2$.
 \end{itemize}
 
 \item The~plots of~the~constancy hypersurfaces of~the~quantities related to~the~electromagnetic field present $A_u$ from $-50$ to~$50$ and~$q$ from $0$ to~$10$, with both ranges divided into $100$ steps.
\end{itemize}

\section{Einstein gravity}
\label{sec:ein}

\begin{figure}[tbp]
\centering
\includegraphics[width=0.8\textwidth]{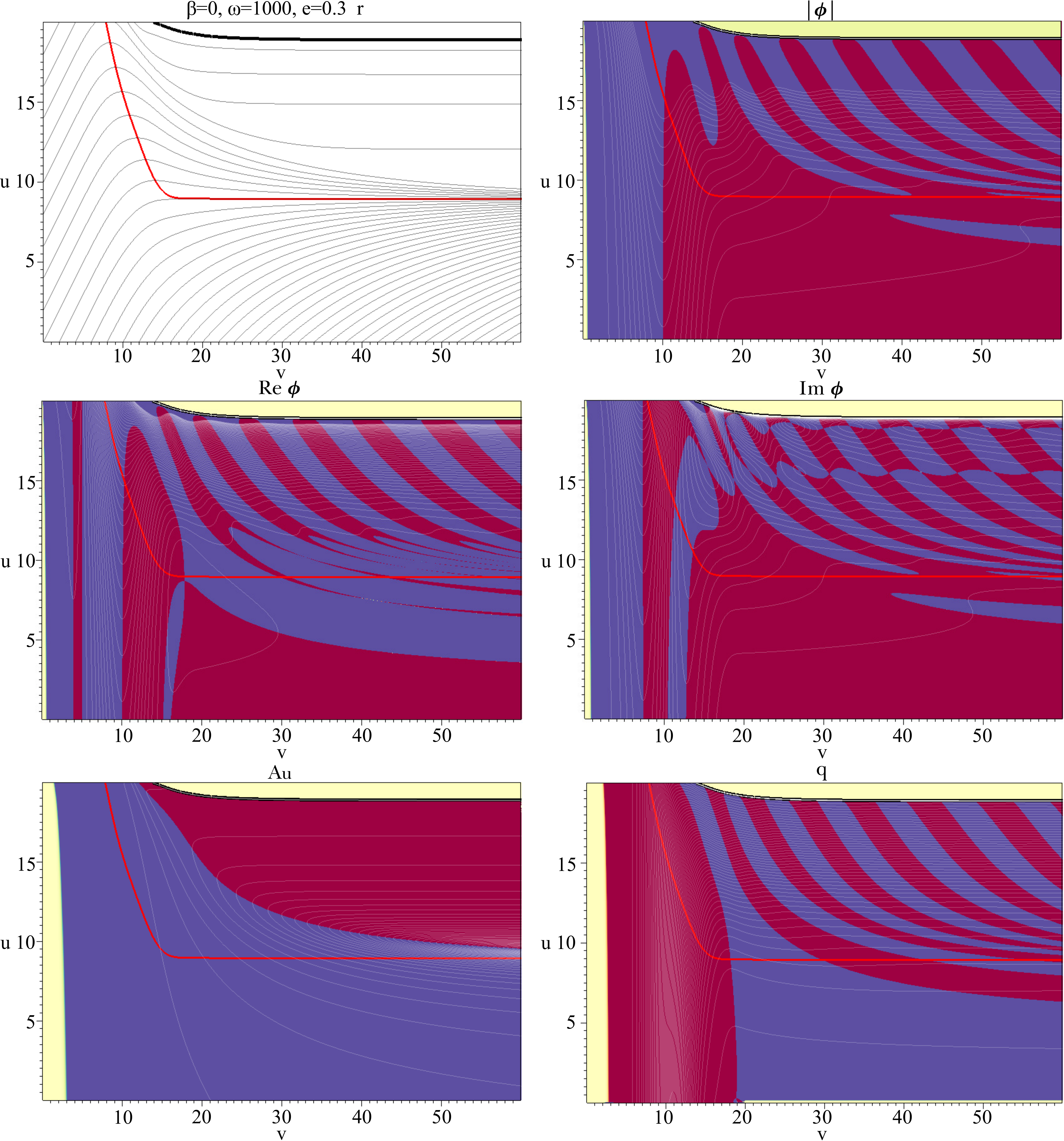}
\caption{(color online) The~diagram of~$r=const.$ lines in~the~$(vu)$-plane for spacetimes formed during the~gravitational evolution of~an~electrically charged scalar field in~the~Einstein theory. The~constancy hypersurfaces of~the~modulus, the~real and~imaginary parts of~the~electrically charged scalar field function ($|\phi|$, $\textrm{Re}\;\phi$ and~$\textrm{Im}\;\phi$, respectively), as~well as~the~$u$-component of~the~$U(1)$ gauge field four-potential ($A_u$) and~the~charge function ($q$).}
\label{fig:Ein}
\end{figure}

The~collapse of~a~self-gravitating electrically charged scalar field within the~Einstein theory was realized by~setting the~evolution parameters $\beta$ and~$\omega$ as~equal to~$0$ and~$1000$, respectively, because the~Einstein limit of~the~Brans-Dicke theory is obtained for $\omega\to\infty$. It~was also checked that the~results are consistent with the~ones obtained when the~Brans-Dicke field is absent, i.e.,~the~amplitude of~its initial profile vanishes and~the~corresponding constants $\beta$ and~$\omega$ are equal to~zero. The~structure of~a~spacetime formed during the~process of~interest, as~well as~the~plots of~constant hypersurfaces of~the~modulus, the~real and~imaginary parts of~the~evolving field function, together with the~$u$-component of~the~$U(1)$ gauge field four-potential and~the~charge function, are presented in~figure~\ref{fig:Ein}.

\subsection{Spacetime structure}
\label{sec:ein-structures}

The~formed spacetime contains a~spacelike central singularity along $r=0$ surrounded by~a~single apparent horizon at~the~hypersurface $r\Pv=0$. The~horizon is spacelike for small values of~advanced time, indicating the~dynamical region of~the~spacetime, and~becomes null as~$v\to\infty$. It~is situated along $u=const.$ hypersurface there and~specifies the~location of~the~event horizon in~the~spacetime, which remains in~its stationary state after the~dynamical collapse. The~lines of~constant $r$ settle along null $u=const.$ hypersurfaces inside the~apparent horizon. It~means that there exists a~Cauchy horizon in~the~spacetime, which is located at~$v=\infty$ null hypersurface inside the~event horizon (for clarification, see figure~\ref{fig:cd-rn}). It~is located outside of~the~computational domain, whose ranges in~both null directions are finite. The~obtained outcome is consistent with the~results of~previous investigations of~the~dynamical spacetimes emerging from the~collapse of~an~electrically charged scalar field~\cite{OrenPiran2003-044013,HongHwangStewartYeom2010-045014,HwangYeom2011-064020,HwangKimYeom2012-055003}. The~described structure is also formed during the~process running in~the~presence of~dark matter~\cite{NakoniecznaRogatkoNakonieczny2015-012}.

\subsection{Dynamical quantities in~the~evolving spacetime}
\label{sec:ein-hypersurfaces}

The~dynamics of~the~complex scalar field is reflected in~the~behavior of~the~constancy hypersurfaces of~the~real and~imaginary parts, as~well as~the~modulus of~the~field function. Their values change most considerably in~the~dynamical region of~the~spacetime and~beyond the~horizon. Nearby the~singularity of~the~former area, the~constancy hypersurfaces of~all the~quantities characterizing the~complex scalar field are spacelike and~their changes are monotonic. Hence, they can be used as~time measurers there. This result agrees with the~conclusion for a~self-gravitating neutral scalar field~\cite{NakoniecznaLewandowski2015-064031}. However, in~contrast to~the~uncharged scalar field, the~electrically charged one cannot serve as~a~`clock' in~the~asymptotic spacetime region of~high curvature. The~constancy hypersurfaces of~the~real and~imaginary parts and~the~modulus of~the~complex field function clearly have either spacelike or~timelike character in~the~vicinity of~the~central singularity at~large values of~$v$.

The~dynamics of~the~$u$-component of~the~Maxwell field four-potential is the~biggest near the~apparent horizon at~large values of~advanced time, while the~dynamics of~the~charge function is the~same as~in~the~case of~the~above-mentioned characteristics of~the~complex scalar field. The~constancy hypersurfaces of~the~quantity $A_u$ are timelike along the~whole singularity. For~this reason, it~cannot be used to~quantify time there. Similarly to~the~quantities $|\phi|$, $\textrm{Re}\;\phi$ and~$\textrm{Im}\;\phi$ described above, the~function $q$ can be employed as~a~`clock' in~the~dynamical spacetime region of~high curvature, because its constancy hypersurfaces are spacelike there and~the~monotonicicty of~their parametrization is preserved. It~is not a~good candidate for a~time measurer nearby the~singularity in~the~asymptotic region, as~its constancy hypersurfaces change their character between spacelike and~timelike.

The~Cauchy horizon is situated at~the~null hypersurface $v=\infty$ inside the~event horizon. Although it~exists outside of~the~computational domain, some conclusions can be drawn about using the~dynamical quantities to~measure time in~its neighborhood, which is also a~region of~high curvature due to~the~mass inflation effect~\cite{PoissonIsrael1990-1796,HodPiran1998-024017,HodPiran1998-024018,HodPiran1998-024019,HodPiran1998-1554,HodPiran1998-1555,HwangLeeYeom2011-006,HansenYeom2015-019}. In~the~region of~large values of~$v$ beyond the~event horizon, the~constancy hypersurfaces of~$\textrm{Re}\;\phi$, $\textrm{Im}\;\phi$ and~$|\phi|$ change their character between spacelike and~timelike as~$u$ changes. The~$u$-component of~the~$U(1)$ gauge field four-potential is almost entirely timelike in~the~region of~interest and~the~behavior of~the~charge function is similar to~the~behavior of~the~quantities characterizing the~complex scalar field. For~the~above reasons, none of~the~dynamical quantities can be used as~a~`clock' in~the~vicinity the~Cauchy horizon.

\section{Brans-Dicke theory}
\label{sec:bd}

\subsection{Spacetime structures}
\label{sec:bd-structures}

\begin{figure}[tbp]
\centering
\includegraphics[width=0.8\textwidth]{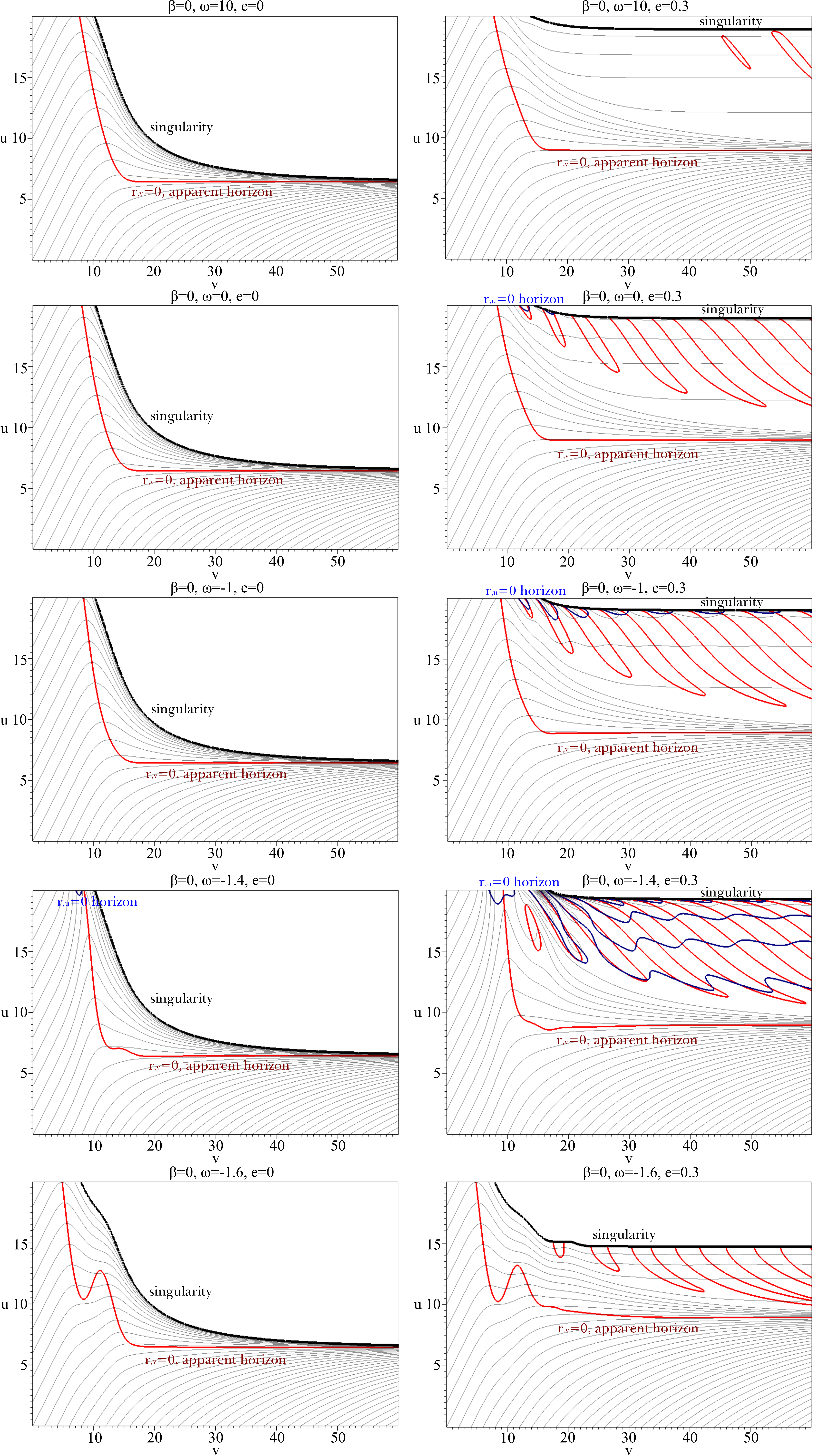}
\caption{(color online) The~diagrams of~$r=const.$ lines in~the~$(vu)$-plane for spacetimes formed during evolutions of~a~neutral complex scalar field (left column) and~electrically charged scalar field (right column) in~the~Brans-Dicke theory for $\beta=0$.}
\label{fig:beta0-r}
\end{figure}

\begin{figure}[tbp]
\centering
\includegraphics[width=0.8\textwidth]{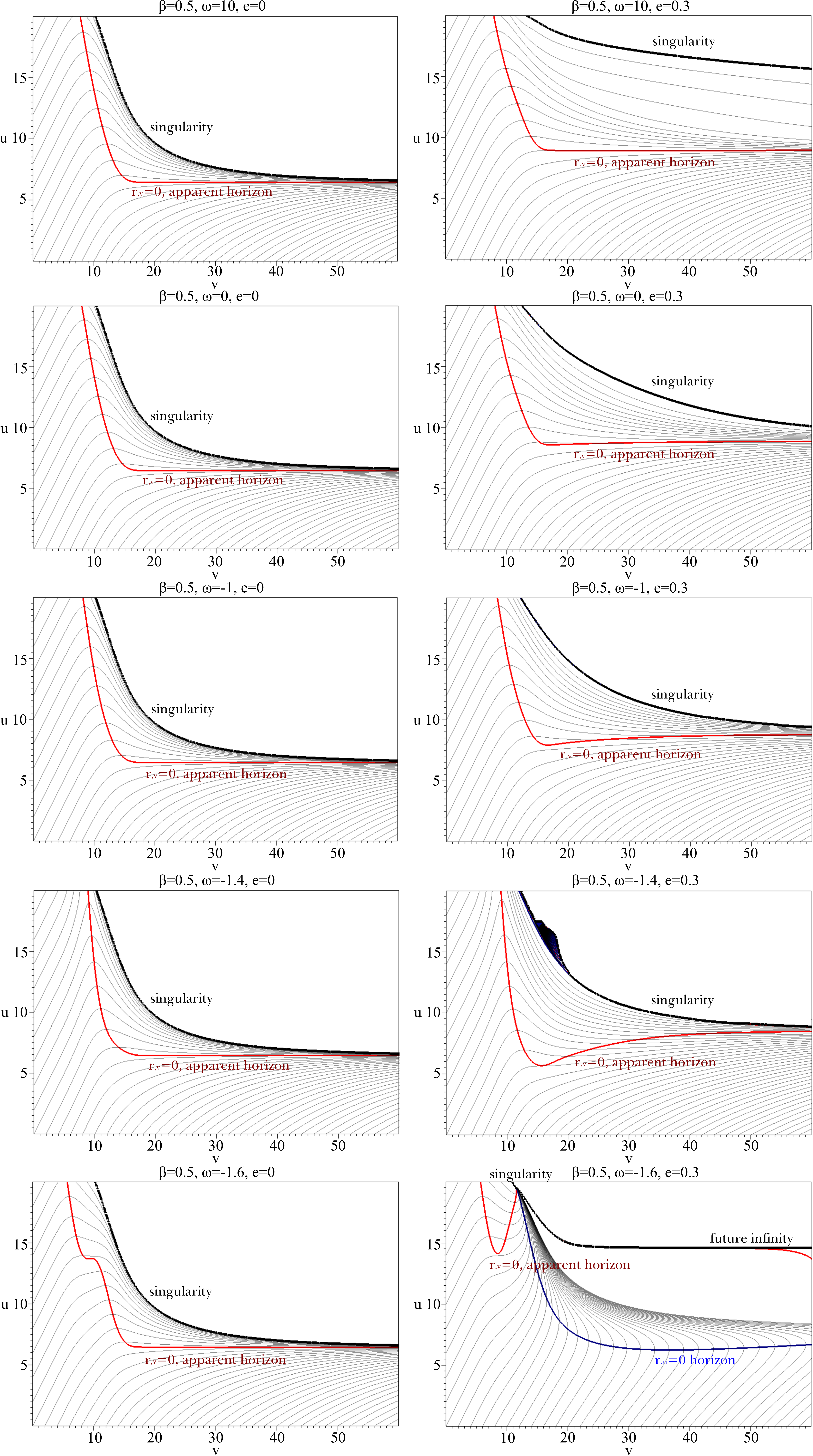}
\caption{(color online) The~diagrams of~$r=const.$ lines in~the~$(vu)$-plane for spacetimes formed during evolutions of~a~neutral complex scalar field (left column) and~electrically charged scalar field (right column) in~the~Brans-Dicke theory for $\beta=0.5$.}
\label{fig:beta05-r}
\end{figure}

\begin{figure}[tbp]
\centering
\includegraphics[width=0.8\textwidth]{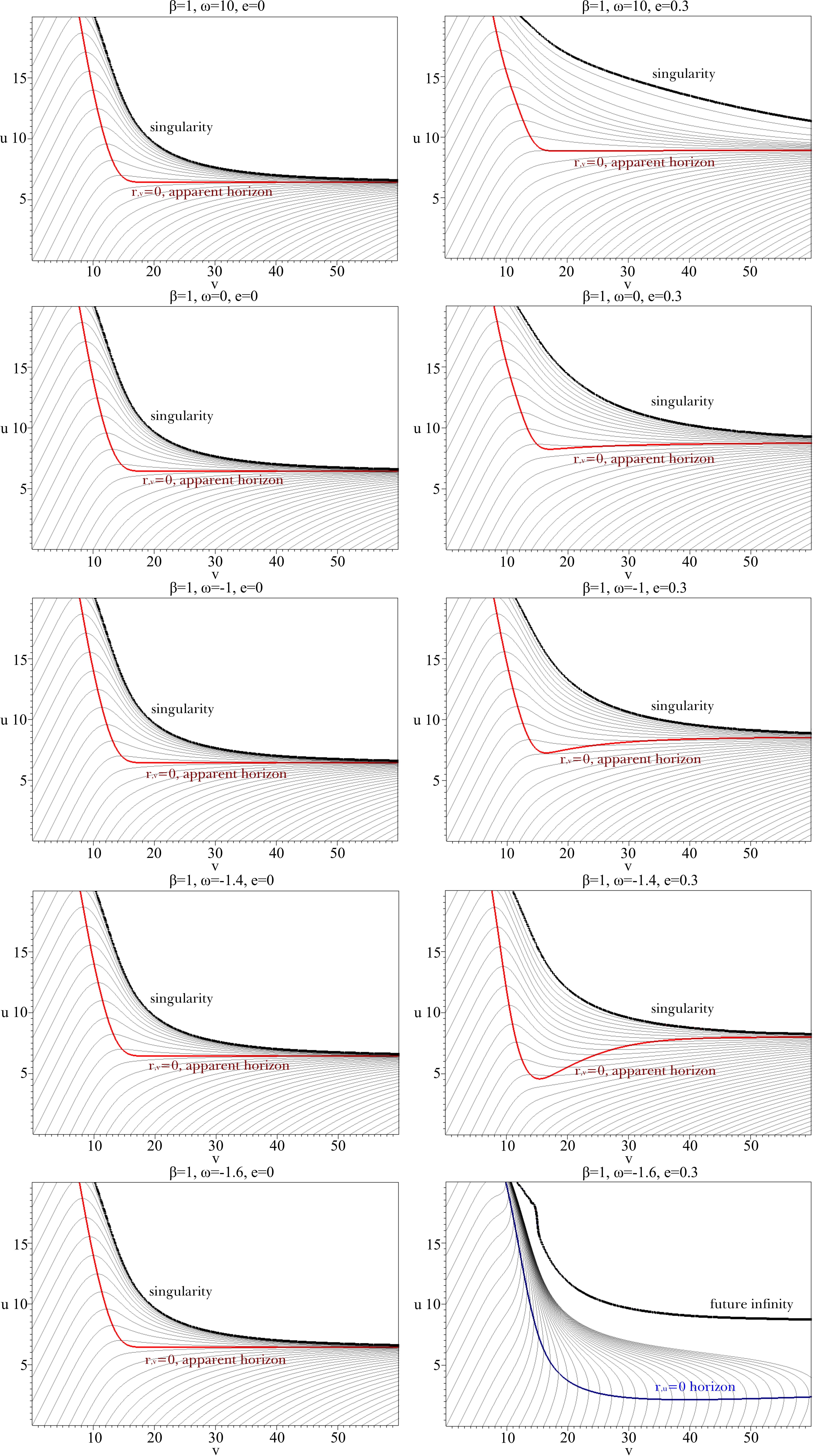}
\caption{(color online) The~diagrams of~$r=const.$ lines in~the~$(vu)$-plane for spacetimes formed during evolutions of~a~neutral complex scalar field (left column) and~electrically charged scalar field (right column) in~the~Brans-Dicke theory for $\beta=1$.}
\label{fig:beta1-r}
\end{figure}

The~structures of~spacetimes resulting from the~gravitational collapse of~an~electrically charged scalar field in~the~Brans-Dicke theory for $\beta$ equal to~$0$, $0.5$ and~$1$ are shown in~figures~\ref{fig:beta0-r},~\ref{fig:beta05-r} and~\ref{fig:beta1-r}, respectively.

\subsubsection{Uncoupled Brans-Dicke and~electrically charged scalar fields}
\label{sec:bd-structures-beta0}

When $\beta$ equals $0$, each spacetime emerging from the~studied collapse contains a~spacelike singularity along $r=0$, wholly surrounded by~an~apparent horizon $r\Pv=0$. The~horizon is spacelike in~the~dynamical region of~the~spacetime and~becomes null and~coincides with the~event horizon when the~spacetime settles in~its stationary state as~$v\to\infty$. For~$\omega$ equal to~$-1.4$ and~$-1.6$, between the~two stages there also exists a~$v$-range within which the~horizon is timelike. For~all values of~the~Brans-Dicke coupling constant, additional $r\Pv=0$ apparent horizons are visible nearby the~singularity. When $\omega$ equals $0$, $-1$ and~$-1.4$, also the~hypersurfaces of~the~apparent horizons $r\Pu=0$ exist in~the~spacetime for large retarded times. In~all the~forming spacetimes, Cauchy horizons are present at~$v=\infty$, what can be inferred from the~fact that beyond the~event horizon the~$r=const.$ lines settle along null $u=const.$ hypersurfaces at~large advanced times. This fact, together with the~existence of~multiple apparent horizons at~large $u$, distinguishes the~collapse with $e\neq 0$ from the~case of~the~vanishing $e$, which produces a~typical Schwarzschild-type spacetime. Such a~spacetime contains a~central spacelike singularity surrounded by~a~single apparent horizon $r\Pv=0$, without a~Cauchy horizon.

\subsubsection{Coupled Brans-Dicke and~electrically charged scalar fields}
\label{sec:bd-structures-beta051}

The~spacetimes formed during the~collapse of~interest for $\beta$ equal to~$0.5$ and~$1$ share many properties. For~$\omega\geqslant -1.4$, in~each of~them there exists a~central spacelike singularity at~$r=0$ surrounded by~a~single apparent horizon $r\Pv=0$. The~horizon is spacelike in~the~dynamical part of~the~spacetime and~tends towards a~null direction as~$v$ increases, indicating the~location of~an~event horizon. When the~Brans-Dicke coupling constant is equal to~$-1$ and~$-1.4$, an~intermediate part of~the~horizon hypersurface is timelike. The~Cauchy horizon is absent in~all spacetimes, analogously to~the~case with $e=0$. A~tendency towards the~Cauchy horizon formation can be seen for the~Brans-Dicke coupling equal to~$10$, as~this large $\omega$ case signifies an~approach to the~Einstein limit of~the~theory.

A~brief comment is required regarding a~bump on~the~singularity in~the~case of~$\beta=0.5$, $\omega=-1.4$, $e=0.3$. This region, which appears as~a~black area on~the~respective plot, covers the~spacetime very close to~the~singularity, which remains spacelike, just as~in the~rest of~the~spacetime. The~fields are highly dynamical there and~many horizons seem to~be folded. Since it~is an~area of~extremely high curvature, its physical meaning is limited for the~current studies, which are conducted within the~scope of~the~classical theory of~gravity.

Entirely distinct structures are observed for $\beta\neq 0$ and~$\omega=-1.6$. When $\beta$ equals~$0.5$, the~central spacelike singularity does not extend to~large values of~advanced time. It~is surrounded by~the~$r\Pv=0$ apparent horizon, which is spacelike and~then timelike for larger values of~$v$. At the~point of~their coincidence another apparent horizon, i.e.,~$r\Pu=0$ appears and~extends to~infinity, initially in~a~spacelike, and~then in~a~null direction. The~future infinity is situated beyond it. For~$\beta=1$, only the~$r\Pu=0$ apparent horizon, which is spacelike in~the~dynamical spacetime region and~becomes null as~$v\to\infty$, is visible in~the~spacetime. Again, the~future infinity is located within it. Such exotic structures were observed previously~\cite{NakoniecznaRogatkoModerski2012-044043,HansenYeom2014-040} and~they can be formed during the~collapse with $\omega=-1.6$, because in~the~ghost limit the~weak cosmic censorship can be violated.

\subsubsection{Overall dependence on~evolution parameters}
\label{sec:bd-structures-sum}

\begin{figure}[tbp]
\subfigure[][]{\includegraphics[width=0.175\textwidth]{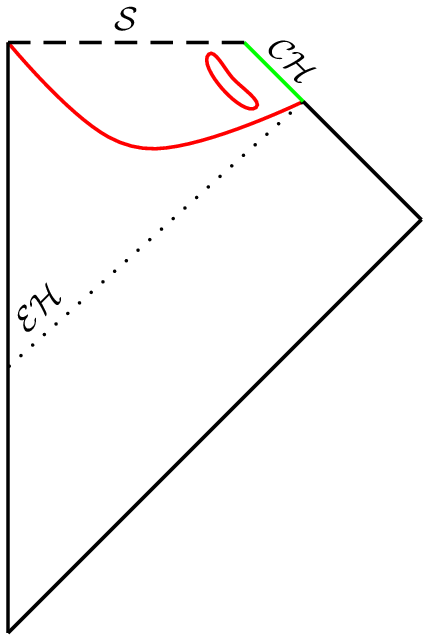}}
\hfill
\subfigure[][]{\includegraphics[width=0.175\textwidth]{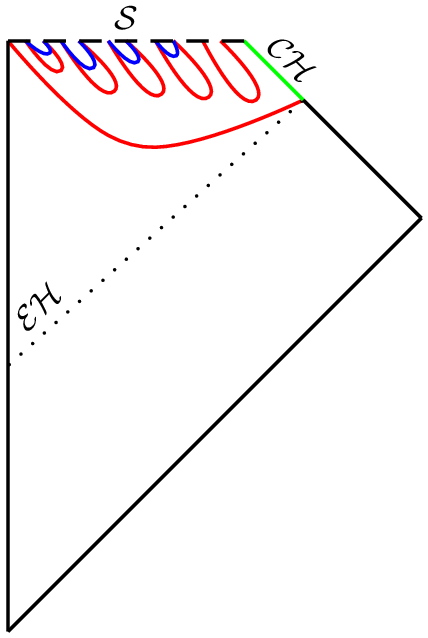}}
\hfill
\subfigure[][]{\includegraphics[width=0.175\textwidth]{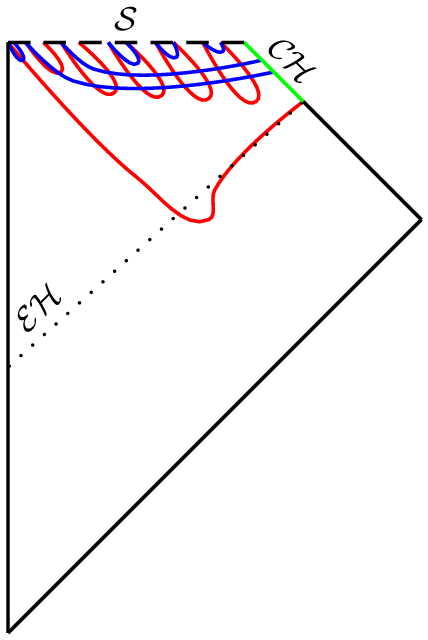}}
\hfill
\subfigure[][]{\includegraphics[width=0.175\textwidth]{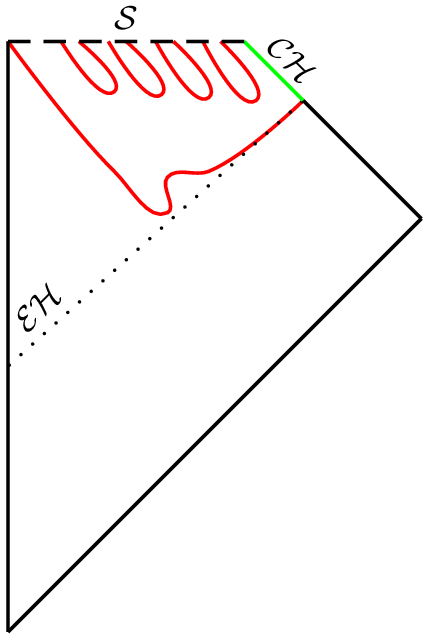}}\\
\subfigure[][]{\includegraphics[width=0.175\textwidth]{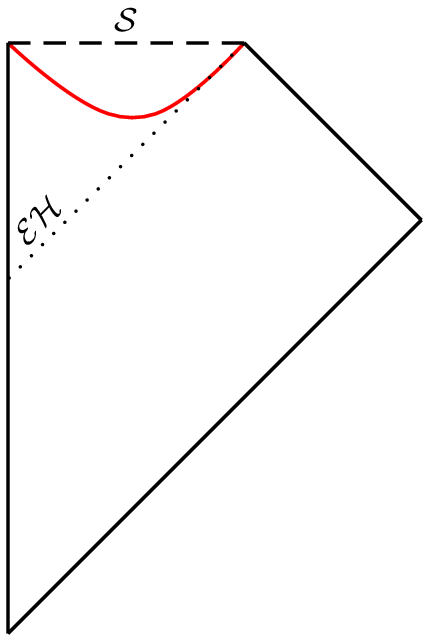}}
\hfill
\subfigure[][]{\includegraphics[width=0.175\textwidth]{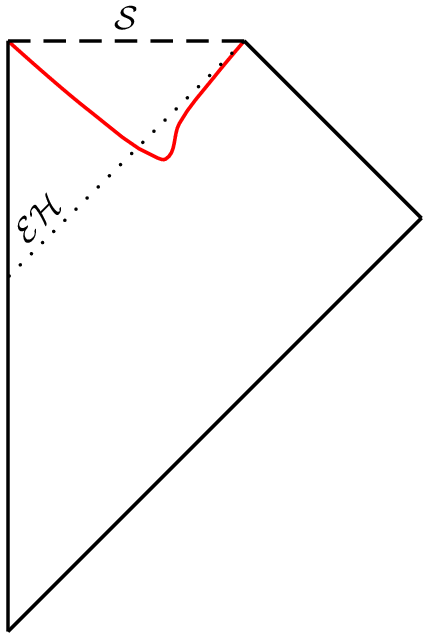}}
\hfill
\subfigure[][]{\includegraphics[width=0.175\textwidth]{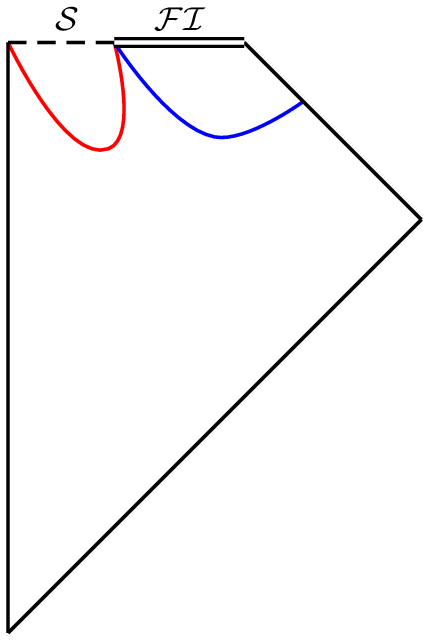}}
\hfill
\subfigure[][]{\includegraphics[width=0.175\textwidth]{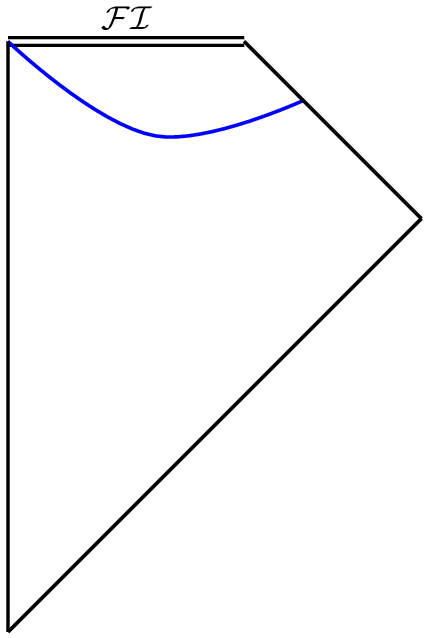}}
\caption{(color online) The~Carter-Penrose diagrams of~spacetimes formed during evolutions of~a~scalar field in~the~Brans-Dicke theory, for the~following sets of~evolution parameters: (a)~$\beta=0$, $\omega=10$, (b)~$\beta=0$, $\omega=0,-1$, (c)~$\beta=0$, $\omega=-1.4$, (d)~$\beta=0$, $\omega=-1.6$, (e)~$\beta\neq 0$, $\omega\geqslant 0$, (f)~$\beta\neq 0$, $\omega=-1,-1.4$, (g)~$\beta=0.5$, $\omega=-1.6$ and~(h)~$\beta=1$, $\omega=-1.6$. The~central singularity along $r=0$, the~event and~Cauchy horizons, as~well as~the~future infinity are denoted as~$\mathcal{S}$, $\mathcal{EH}$, $\mathcal{CH}$ and~$\mathcal{FI}$, respectively.}
\label{fig:CPdiags}
\end{figure}

Similarly to~the~case of~an~uncharged collapse studied in~paper~I, the~variety and~complexity of~the~emerging spacetime structures decrease as~$\beta$ increases. The~exotic structures formed for $\omega=-1.6$ and~$\beta\neq 0$ constitute an~exception in~this regard. It~is worth emphasizing that in~none of~the~cases the~collapse proceeds similarly to~the~process running in~the~Einstein gravity, whose outcome was described in~section~\ref{sec:ein-structures}. The~Cauchy horizon forms when the~Brans-Dicke field is not coupled with the~complex scalar one and~its emergence is prevented for the~remaining values of~$\beta$. The~existence of~the~Cauchy horizon is always accompanied by~the~presence of~multiple apparent horizon hypersurfaces within the~event horizon. Negative values of~the~parameter $\omega$ favor an~existence of~a~timelike part of~the~apparent horizon $r\Pv=0$ between its spacelike and~null sections. A~summary of~causal structures of~spacetimes obtained as~a~result of~the~examined process is presented in~figure~\ref{fig:CPdiags} in~the~form of~Carter-Penrose diagrams.

\subsection{Dynamical quantities in~evolving spacetimes}
\label{sec:bd-hypersurfaces}

The~evolution of~the~Brans-Dicke and~electrically charged scalar fields~\eqref{eqn:BD}--\eqref{eqn:sf1} in~double null coordinates is governed by~the~following equations:
\ben\label{eqn:BDfield-eqn2}
\Phi\Puv &=& -\frac{fZ+gW}{r} + \frac{\Phi^\beta \beta\, q^2\alpha^2}{2r^4\left(3+2\omega\right)} 
- \frac{\Phi^\beta\left(1-\beta\right)}{3+2\omega} \big[w\bz+z\bw+ieA_u\left(\bz s-z\bs\right)\big], \\
\label{eqn:field-eqn2}
\phi\Puv &=& -\frac{fz+gw}{r} - ieA_uz - \frac{ieA_ugs}{r} - \frac{ie}{4r^2}\alpha^2 qs 
- \frac{\beta}{2\Phi} \left( Wz + Zw + iesA_uZ \right).
\een
The~evolution of~the~$u$-component of~the~$U(1)$ gauge field four-potential and~the~charge function in~the~$(vu)$-plane is governed by~equations~\eqref{eqn:Au-dn} and~\eqref{eqn:q-dn}, respectively. As~may be inferred from~\eqref{eqn:BDfield-eqn2}, the~case of~$\omega=-1.5$ is a~singular point of~the~evolution equation.

\subsubsection{Type IIA~model}
\label{sec:bd-dyn-beta0}

\begin{figure}[tbp]
\centering
\includegraphics[width=0.8\textwidth]{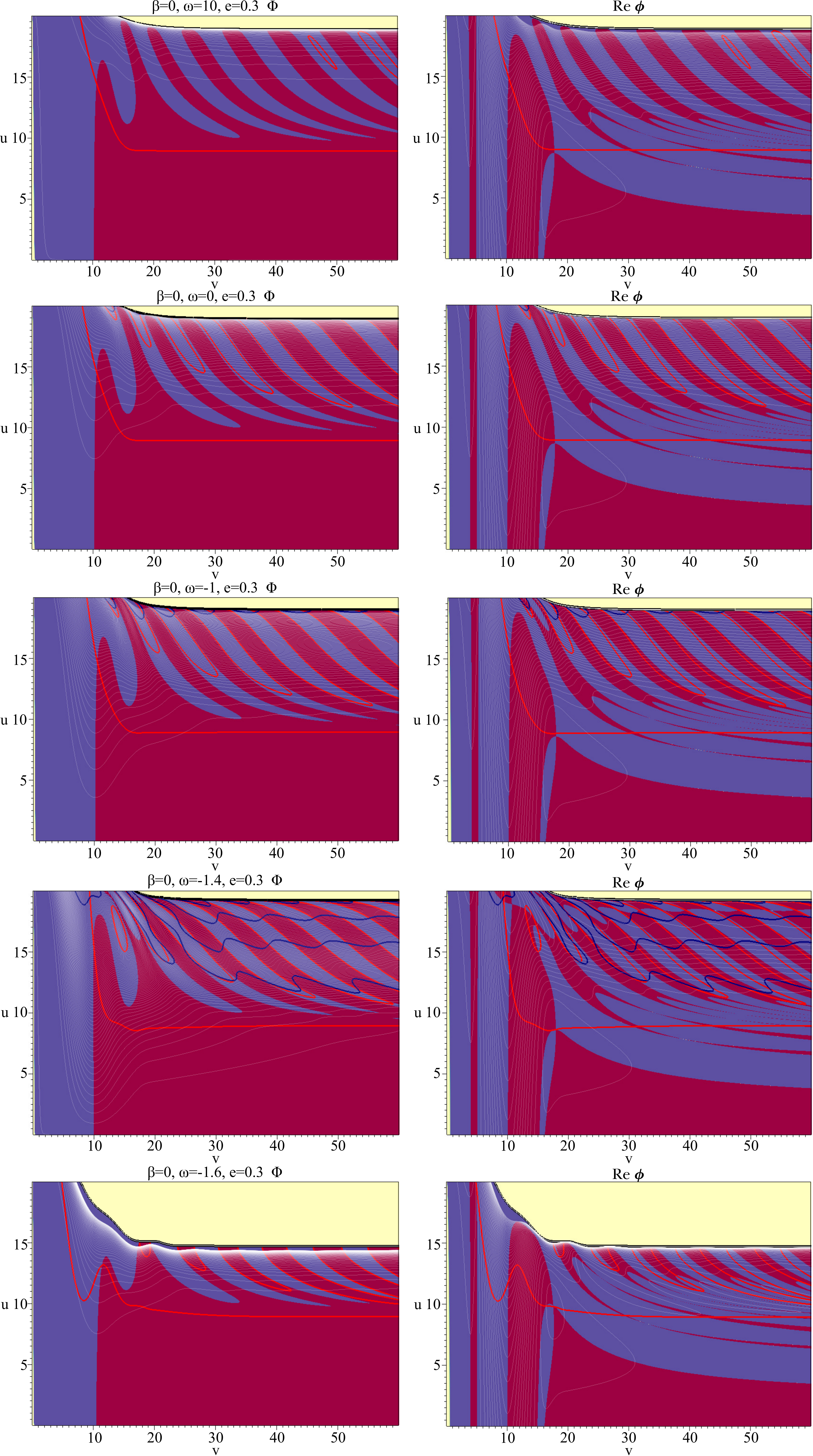}
\caption{(color online) The~constancy hypersurfaces of~the~Brans-Dicke field (left column) and~the~real part of~the~electrically charged scalar field function (right column) for evolutions conducted within the~Brans-Dicke theory for $\beta=0$.}
\label{fig:beta0-ch}
\end{figure}

The~hypersurfaces of~constant values of~the~Brans-Dicke field and~the~real part of~the~electrically charged scalar field for the~evolutions proceeding with $\beta=0$ are shown in~figure~\ref{fig:beta0-ch}. The~Brans-Dicke and~the~electrically charged fields become more dynamical as~the~singularity is approached except the~latter when $\omega$ equals $-1.4$, for which the~opposite tendency is observed. The~Brans-Dicke field varies more substantially in~the~dynamical region of~spacetime as~$\omega$ decreases, except the~case of~$\omega=-1.6$, in~which the~changes of~the~field values are less considerable in~comparison to, e.g.,~$\omega=-1.4$. The~Brans-Dicke parameter does not influence the~tendencies in~the~dynamics of~the~electrically charged scalar field, whose values vary in~a~similar manner within the~dynamical area for all the~studied values of~$\omega$.

In~all the~emerging spacetimes the~constancy hypersurfaces of~the~Brans-Dicke field are spacelike nearby the~central singularity in~its dynamical part. It~means that the~field is a~good candidate for a~time measurer in~this area, also because its values change monotonically there. For~larger values of~advanced time, the~character of~the~hypersurfaces changes in~the~vicinity of~the~singularity and~they are either spacelike or~timelike there. Hence, the~Brans-Dicke field cannot be used a~`clock' in~the~asymptotic spacetime region. This result is in~general consistent with the~outcomes obtained in~paper~I for the~uncharged case. The~only difference is that in~the~dynamical area, the~Brans-Dicke field can be employed to~quantify time further away from the~singularity, as~in the~current case there are none nonspacelike hypersurfaces reaching $r=0$, even at~single points.

The~constancy hypersurfaces of~the~real part of~the~electrically charged scalar field are spacelike in~the~dynamical spacetime region of~high curvature. They also change monotonically there and~thus, the~field is a~potential time measurer in~this area. On~the~contrary, similarly to~the~above-mentioned case of~the~Brans-Dicke field, when $v$ increases to~the~values at~which the~apparent horizon settles along a~null hypersurface, the~constancy hypersurfaces are either spacelike or~timelike near the~singularity. For~this reason, the~electrically charged scalar field cannot play a~role of~a~`clock' there. The~conclusions are in~agreement with the~neutral case of~paper~I.

\begin{figure}[tbp]
\centering
\includegraphics[width=0.8\textwidth]{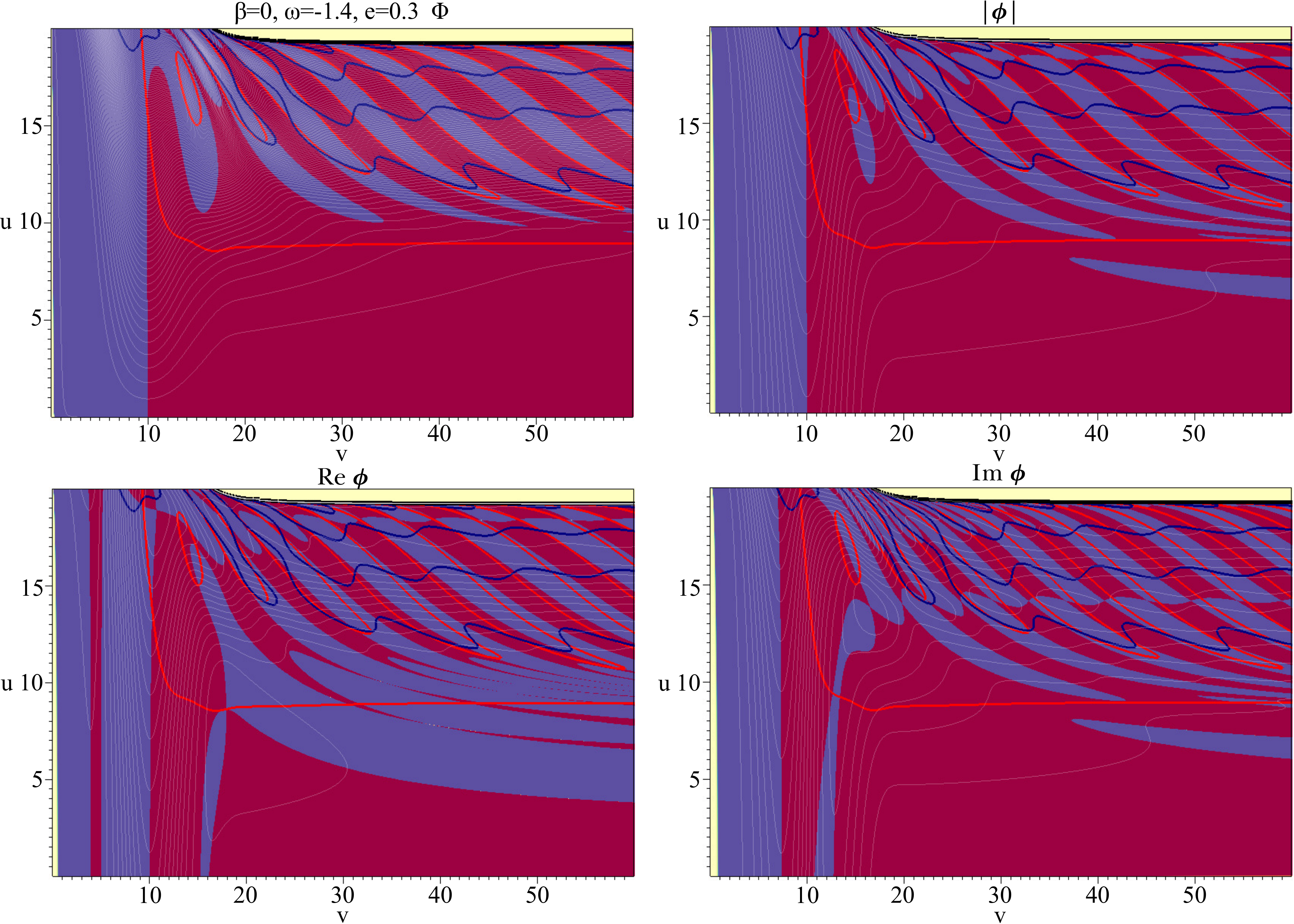}
\caption{(color online) The~constancy hypersurfaces of~the~Brans-Dicke field $\Phi$, as~well as~the~modulus, the~real and~imaginary parts of~the~electrically charged scalar field function ($|\phi|$, $\textrm{Re}\;\phi$ and~$\textrm{Im}\;\phi$, respectively) for the~evolution described by~the~parameters $\beta=0$ and~$\omega=-1.4$.}
\label{fig:beta0-ch-special}
\end{figure}

The~constancy hypersurfaces of~the~Brans-Dicke field, as~well as~the~modulus, the~real and~imaginary parts of~the~electrically charged scalar field function for the~sample evolution with $\beta=0$ and~$\omega=-1.4$ are shown in~figure~\ref{fig:beta0-ch-special}. The~behavior of~the~constancy hypersurfaces of~the~imaginary part of~the~complex field and~its modulus is similar to~their behavior for the~real part of~the~field function. The~field dynamics is most apparent in~the~dynamical spacetime region and~beyond the~event horizon, where it~decreases as~the~singularity is approached. The~constancy hypersurfaces are spacelike and~change monotonically in~the~close neighborhood of~the~singularity for small $v$, and~their character is either spacelike or~timelike as~$v\to\infty$. For~these reasons, all the~characteristics of~the~electrically charged scalar field can be used as~`clocks' in~the~highly curved dynamical area and~are excluded in~this regard for large values of~advanced time.

\begin{figure}[tbp]
\centering
\includegraphics[width=0.8\textwidth]{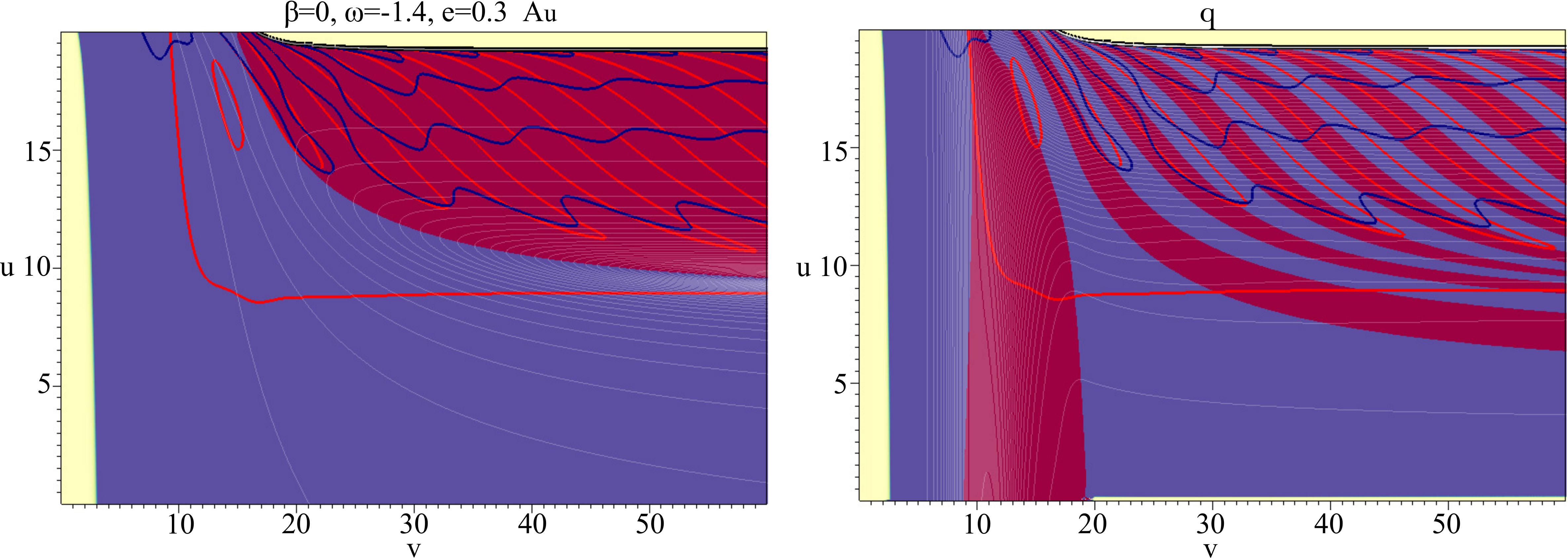}
\caption{(color online) The~constancy hypersurfaces of~the~$u$-component of~the~$U(1)$ gauge field four-potential ($A_u$) and~the~charge function ($q$) for the~evolution described by~the~parameters $\beta=0$ and~$\omega=-1.4$.}
\label{fig:beta0-Auq}
\end{figure}

Since none of~the~above quantities can be used to~measure time along the~whole central singularity at~$r=0$, the~$u$-component of~the~$U(1)$ gauge field four-potential and~the~charge function were also tested in~this respect. Their constancy hypersurfaces for the~sample evolution with $\beta=0$ and~$\omega=-1.4$ are shown in~figure~\ref{fig:beta0-Auq}. The~former is mostly dynamical for large advanced times in~the~vicinity of~the~event horizon. The~constancy hypersurfaces of~$A_u$ are timelike along the~whole singularity, so it~definitely cannot serve as~a~time quantifier in~the~regions of~high curvature during the~collapse. The~values of~the~charge function change considerably in~the~dynamical region of~the~spacetime and~the~dynamics increases for all $v$ when the~singularity is approached. The~nature of~the~constancy hypersurfaces of~$q$ is either spacelike or~timelike nearby the~singularity and~for this reason this quantity is not a~good candidate for a~`clock' during the~examined process.

As~may be inferred from figures~\ref{fig:beta0-ch} and~\ref{fig:beta0-ch-special}, as~the~Cauchy horizon is approached the~nature of~the~hypersurfaces of~constant $\textrm{Re}\;\phi$, $\textrm{Im}\;\phi$ and~$|\phi|$ change between spacelike and~timelike as~$u$ changes. For~this reason, these quantities cannot be used as~time measurers nearby the~Cauchy horizon. As~can be seen in~figure~\ref{fig:beta0-Auq}, the~$u$-component of~the~$U(1)$ gauge field four-potential is timelike in~almost the~whole region of~interest and~the~behavior of~the~charge function is similar to~the~behavior of~the~quantities characterizing the~complex scalar field. Thus, $A_u$ and~$q$ are also excluded as~time quantifiers as~the~Cauchy horizon is approached. This result obtained for the~vicinity of~the~Cauchy horizon is the~same as~in~the~case of~the~collapse proceeding in~the~Einstein gravity, described in~section~\ref{sec:ein-hypersurfaces}.

\subsubsection{Type I and~heterotic models}
\label{sec:bd-dyn-beta051}

\begin{figure}[tbp]
\centering
\includegraphics[width=0.8\textwidth]{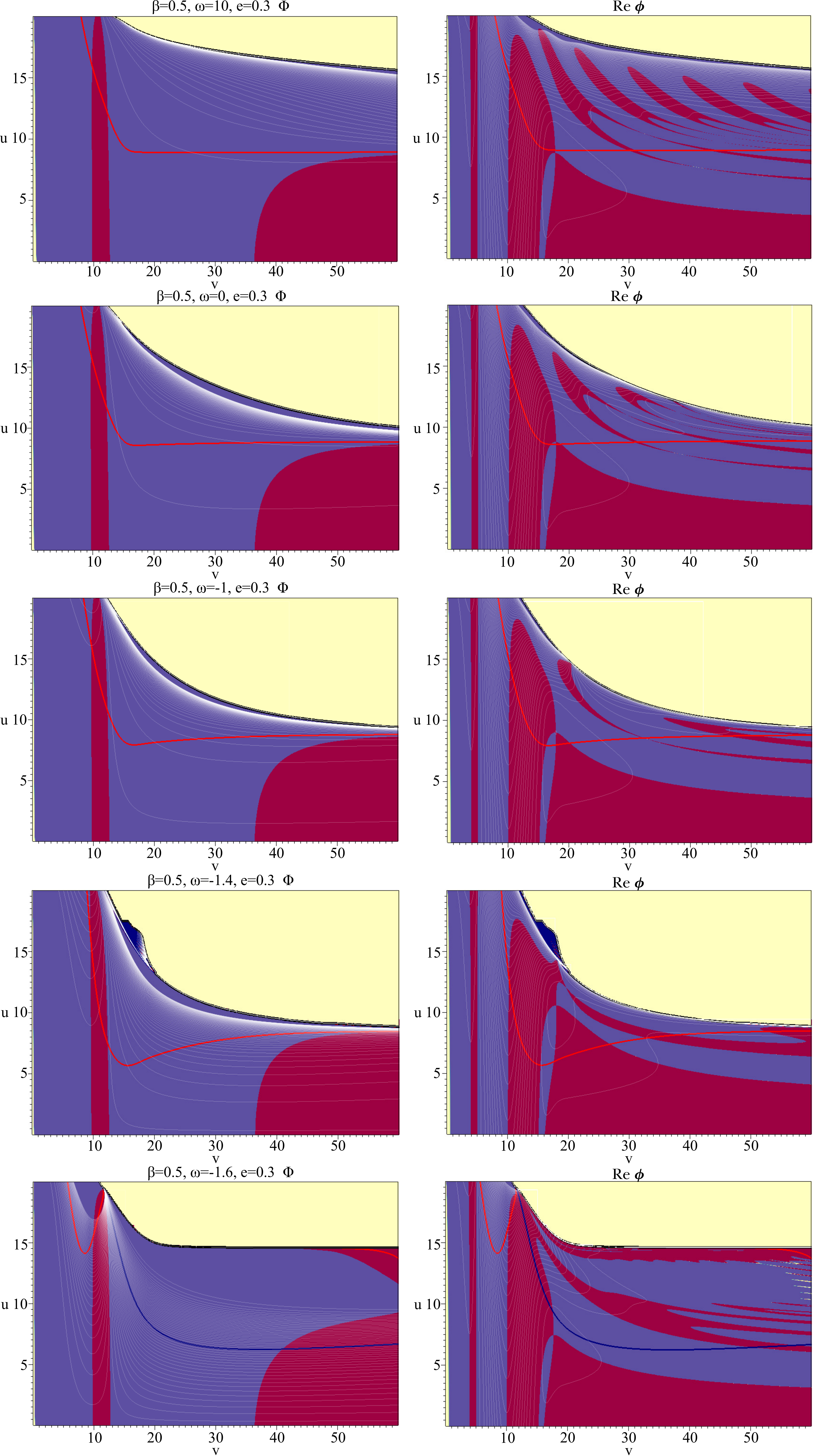}
\caption{(color online) The~constancy hypersurfaces of~the~Brans-Dicke field (left column) and~the~real part of~the~electrically charged scalar field function (right column) for evolutions conducted within the~Brans-Dicke theory for $\beta=0.5$.}
\label{fig:beta05-ch}
\end{figure}

\begin{figure}[tbp]
\centering
\includegraphics[width=0.8\textwidth]{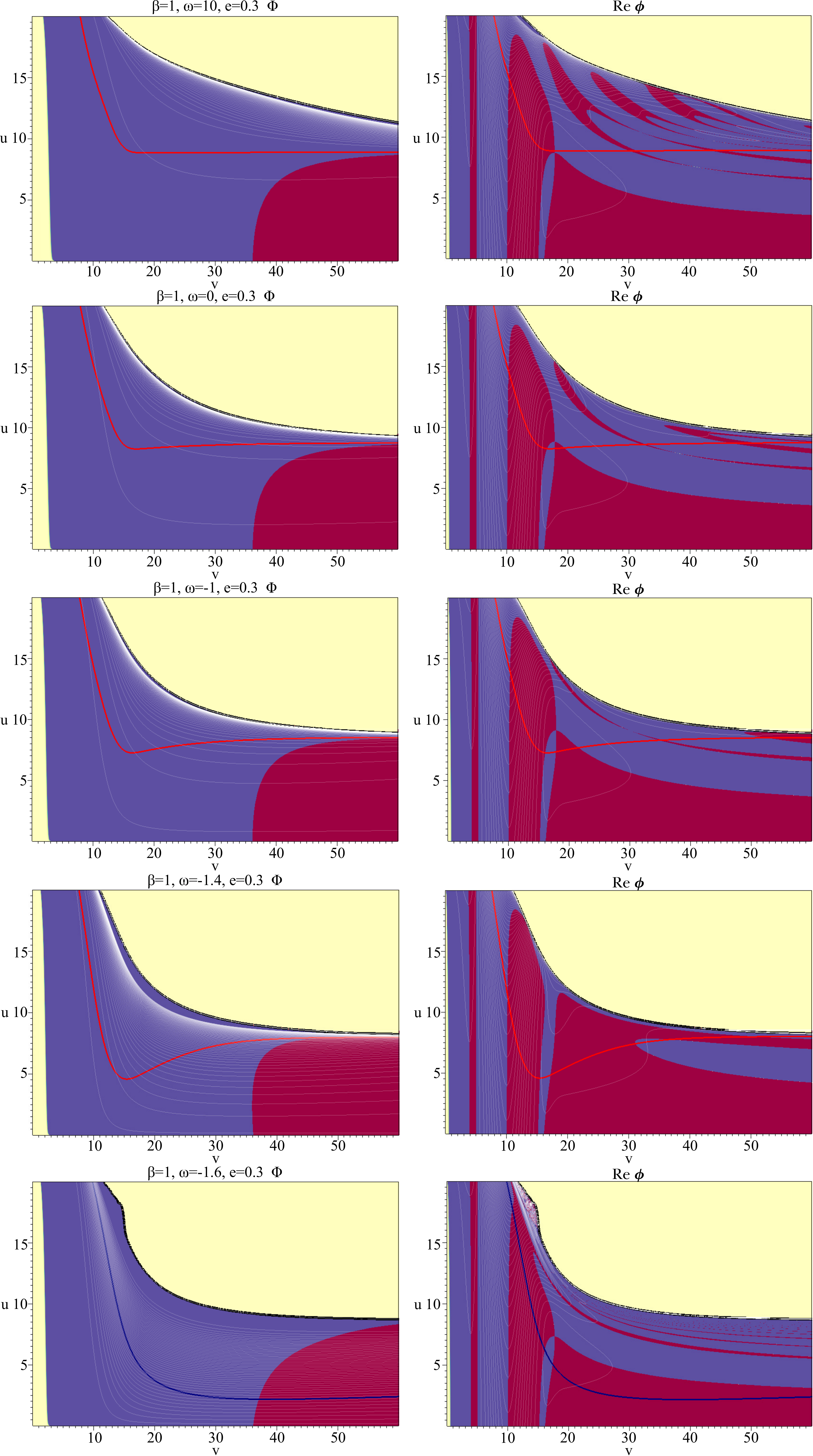}
\caption{(color online) The~constancy hypersurfaces of~the~Brans-Dicke field (left column) and~the~real part of~the~electrically charged scalar field function (right column) for evolutions conducted within the~Brans-Dicke theory for $\beta=1$.}
\label{fig:beta1-ch}
\end{figure}

The~constancy hypersurfaces of~the~Brans-Dicke field and~the~real part of~the~electrically charged scalar field for the~evolutions proceeding with $\beta$ equal to~$0.5$ and~$1$ are shown in~figures~\ref{fig:beta05-ch} and~\ref{fig:beta1-ch}, respectively. The~values of~the~Brans-Dicke field function vary more and~more considerably as~the~central singularity is approached along both null directions. Its~dynamics in~the~dynamical spacetime region is less apparent and~it~increases as~$\omega$ decreases. In~the~case of~$\omega=-1.6$, the~Brans-Dicke field function values change dynamically nearby the~point, at~which the~$r\Pv=0$ and~$r\Pu=0$ apparent horizons meet the~singularity and~the~future infinity. The~electrically charged scalar field is most dynamical within the~$v$-range, in~which the~$r\Pv=0$ apparent horizon is spacelike. The~values of~the~real part of~its field function change significantly in~the~dynamical spacetime region and~this behavior does not depend on~the~parameter $\omega$. The~field becomes more dynamical as~the~singularity is approached and~this phenomenon diminishes as~$\omega$ decreases.

When both scalar fields present in~the~system are coupled, the~constancy hypersurfaces of~the~Brans-Dicke field are spacelike nearby the~whole singularity. The~field function values vary monotonically in~the~region of~high curvature. When $\beta$ equals $0.5$, there can exist a~single point at~the~singularity, at~which a~nonspacelike constancy hypersurface reaches~it. However, as~was explained in~paper~I, such isolated points do not exclude the~field from being a~time quantifier. For~the~above reasons, the~Brans-Dicke field can be employed to~measure time in~the~neighborhood of~the~singular $r=0$. The~above outcomes obtained for the~charged scalar field collapse are different from the~results achieved for the~neutral one in~paper~I. When a~real or~complex uncharged scalar field accompanied the~Brans-Dicke field during the~process, the~latter could be a~time measurer only in~the~dynamical spacetime region when $\beta$ was equal to~$0.5$. Hence, the~existence of~electric charge in~the~spacetime enables the~Brans-Dicke field to~become a~good candidate for a~`clock' in~the~entire vicinity of~the~central singularity for both examined values of~$\beta\neq 0$.

The~hypersurfaces of~constant values of~the~real part of~the~electrically charged scalar field are spacelike along the~singularities in~all cases with a~nonvanishing $\beta$, except $\omega=10$ and~$\beta=1$. There can only exist separated points, at~which a~nonspacelike hypersurface can reach the~singularity. These points, as~was mentioned above, do not prevent the~quantity from being a~`clock' nearby the~singularity. Moreover, the~field function values change monotonically when the~singularity is approached. Hence, the~charged scalar field can be treated as~a~time measurer along the~central singularity for all the~cases apart from $\omega=10$ with $\beta=1$, for which timelike constancy hypersurfaces reach the~singularity at~large values of~advanced time. This exceptional behavior probably signals an~approach of~the~Einstein limit of~the~theory, which appears for large values of~the~Brans-Dicke coupling.

\begin{figure}[tbp]
\centering
\includegraphics[width=0.8\textwidth]{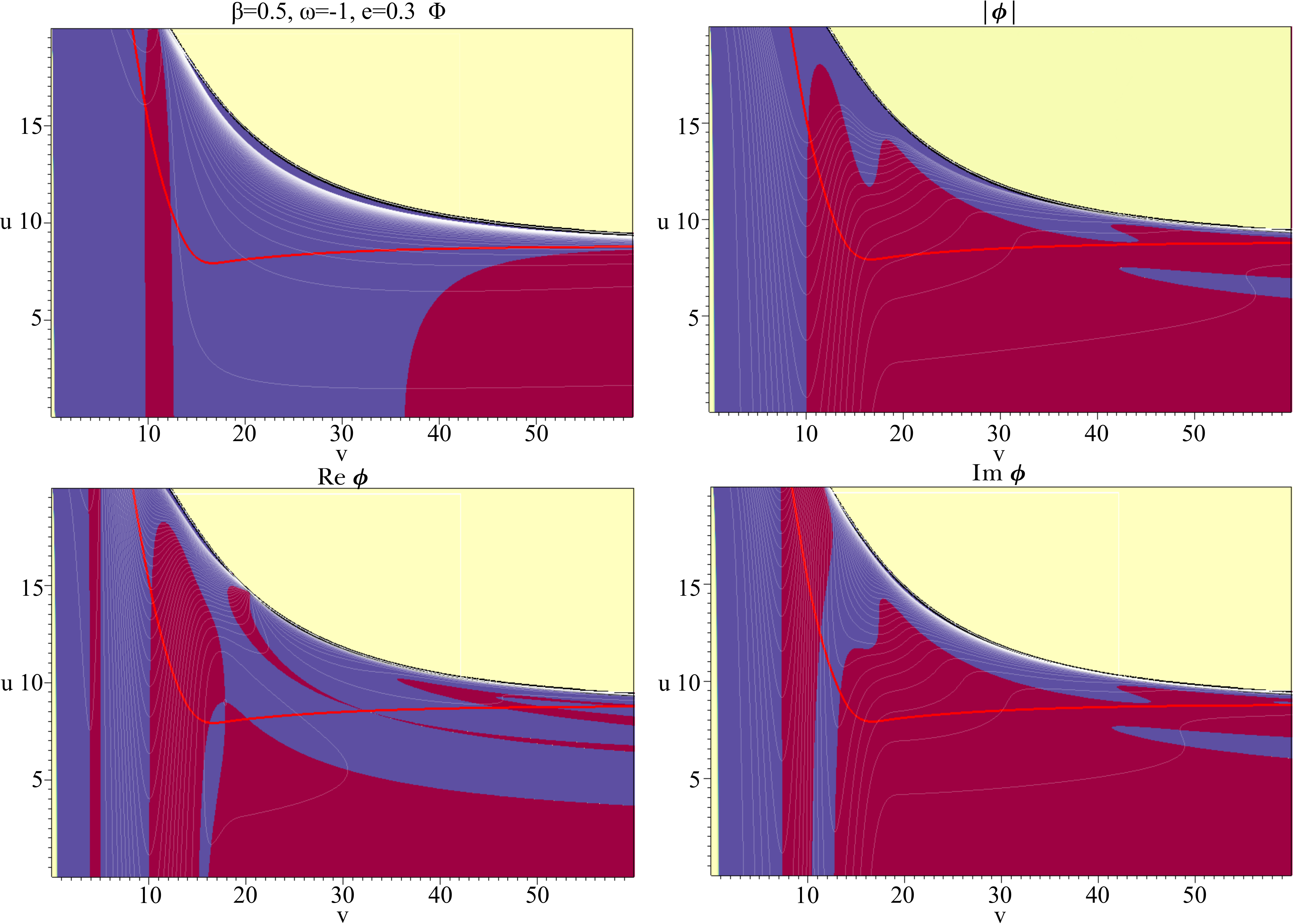}
\caption{(color online) The~constancy hypersurfaces of~the~Brans-Dicke field $\Phi$, as~well as~the~modulus, the~real and~imaginary parts of~the~electrically charged scalar field function ($|\phi|$, $\textrm{Re}\;\phi$ and~$\textrm{Im}\;\phi$, respectively) for the~evolution described by~the~parameters $\beta=0.5$ and~$\omega=-1$.}
\label{fig:beta05-ch-special}
\end{figure}

\begin{figure}[tbp]
\centering
\includegraphics[width=0.8\textwidth]{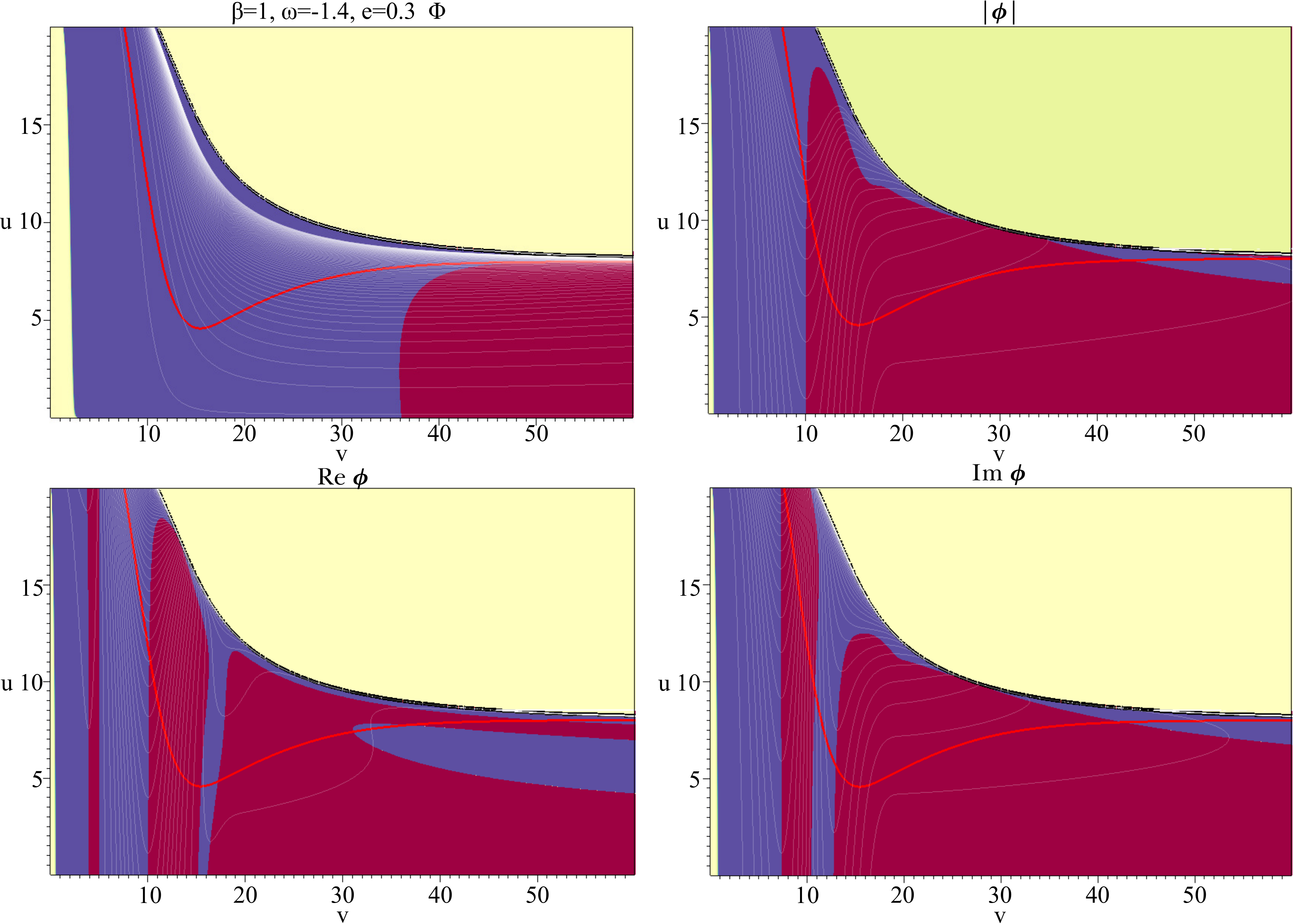}
\caption{(color online) The~constancy hypersurfaces of~the~Brans-Dicke field $\Phi$, as~well as~the~modulus, the~real and~imaginary parts of~the~electrically charged scalar field function ($|\phi|$, $\textrm{Re}\;\phi$ and~$\textrm{Im}\;\phi$, respectively) for the~evolution described by~the~parameters $\beta=1$ and~$\omega=-1.4$.}
\label{fig:beta1-ch-special}
\end{figure}

The~hypersurfaces of~constant values of~the~Brans-Dicke field, as~well as~the~modulus, the~real and~imaginary parts of~the~electrically charged scalar field function for the~evolution with $\beta=0.5$ and~$\omega=-1$ are shown in~figure~\ref{fig:beta05-ch-special}. The~same set of~quantities for the~collapse running for $\beta=1$ and~$\omega=-1.4$ is shown in~figure~\ref{fig:beta1-ch-special}. Both the~imaginary part of~the~field function and~its modulus behave analogously to~the~real part of~the~complex scalar field. They are spacelike nearby the~central singularity and~they change monotonically there. Although there can exist separated points, at~which a~nonspacelike hypersurface reaches the~singularity, both of~the~quantities can be treated as~`clocks' in~the~whole region of~high curvature. Since all the~investigated field characteristics provide a~time measure in~the~area of~interest, there does not exist the~need for examining other quantities in~this respect.

\subsubsection{Overall dependence on~evolution parameters}
\label{sec:bd-dyn-sum}

The~dynamics of~the~Brans-Dicke field is most evident as~the~central singularity is approached for all the~investigated values of~$\beta$ and~$\omega$. The~field function values change more significantly for smaller retarded times as~$\omega$ decreases, with one exception of~$\beta=0$ and~$\omega=-1.6$. The~nature of~the~Brans-Dicke field constancy hypersurfaces strongly depends on~whether the~parameter $\beta$ vanishes or~not. Thus, it~can be suspected that it~is mainly related to~the~second term of~the~right-hand side of~the~equation~\eqref{eqn:BDfield-eqn2}. In~the~latter case, the~hypersurfaces are spacelike in~the~whole region of~high curvature, while in~the~former case they can be timelike nearby the~singularity for large values of~$v$. Whatever the~value of~$\beta$ and~$\omega$, the~real part of~the~electrically charged scalar field is most dynamical in~the~dynamical spacetime region and~its values vary more significantly in~the~vicinity of~the~singularity, apart from $\beta=0$ and~$\omega=-1.4$, in~which the~latter tendency is reversed. Similarly to~the~Brans-Dicke field function, $\textrm{Re}\;\phi$ is always spacelike in~the~neighborhood of~the~singularity when $\beta\neq 0$ and~can be timelike at~large advanced times when $\beta$ vanishes.

\section{Conclusions}
\label{sec:conclusions}

The~dynamical collapse of~a~self-gravitating electrically charged scalar field was examined within the~frameworks of~the~Einstein gravity and~the~Brans-Dicke theory. In~the~latter case, a~set of~values of~the~Brans-Dicke coupling constant and~the~coupling between the~Brans-Dicke and~the~electrically charged scalar fields was considered. The~assumed values of~$\omega$ allowed us to investigate the~large $\omega$, $f(R)$, dilatonic, brane-world and~ghost limits of~the~theory. The~chosen values of~the~$\beta$ parameter were motivated by the~type~IIA, type~I and~heterotic string theory-inspired models.

Apart from the~emerging spacetime structures, which were described in~each of~the~investigated cases, a~possibility of~measuring time with the~use of~dynamical quantities present in~the~system was assessed. The~current paper broadens the~scope of~\cite{NakoniecznaLewandowski2015-064031} and~complements the~analyses described in~paper~I, which referred to the~neutral scalar field collapse in~the~general relativistic and~Brans-Dicke regimes, respectively.

The~outcome of~the~collapse of~an~electrically charged scalar field in~the~Einstein gravity is a~dynamical Reissner-Nordstr{\"o}m spacetime. The~central spacelike singularity along $r=0$ is surrounded by a~single apparent horizon $r\Pv=0$, which settles along the~event horizon of~the~spacetime in~the~region, where $v\to\infty$. The~Cauchy horizon is situated at the~null hypersurface $v=\infty$ beyond the~event horizon.

The~real and~imaginary parts of~the~complex scalar field function, as~well as~its modulus and~the~charge function are most dynamical in~the~spacetime region, in~which the~apparent horizon is spacelike. Their values change more and~more considerably as~the~singularity is approached. The~variations of~the~values of~the~$u$-component of~the~Maxwell field four-potential are most significant nearby the~horizon at large values of~advanced time. Nearby the~singularity, all the~quantities characterizing the~electrically charged scalar field and~the~charge function can be employed as~`clocks' in~the~dynamical spacetime region, because their constancy hypersurfaces are spacelike and~their values change monotonically in~the~area. For~larger values of~$v$, measuring time with their use in~the~region of~high curvature is impossible, because the~character of~the~constancy hypersurfaces changes between spacelike and~timelike. The~quantity $A_u$ cannot be used to quantify time, as~the~hypersurfaces of~its constant values are timelike along the~whole singularity. None of~the~above-mentioned quantities is a~good candidate for a~time measurer in~the~neighborhood of~the~Cauchy horizon, due to the~possible timelike character of~their constancy hypersurfaces as~the~horizon is approached.

In~comparison to the~neutral scalar field collapse described from the~viewpoint of~time quantification using a~dynamical field in~\cite{NakoniecznaLewandowski2015-064031}, the~results obtained in~the~case of~an~electrically charged scalar field case support the~conclusion that the~evolving scalar field has a~potential for measuring time in~the~dynamical spacetime region nearby the~singularity. On~the~contrary, the~existence of~electric charge in~the~spacetime excludes the~possibility of~using the~scalar field as~a~`clock' in~the~asymptotic region of~high curvature, i.e.,~for large values of~advanced time in~the~vicinity of~the~singularity.

During the~collapse proceeding in~the~Brans-Dicke theory, when $\beta$ equals $0$ each of the~emerging spacetimes contains a~central spacelike singularity surrounded by an~apparent horizon $r\Pv=0$, which coincides with the~event horizon as~$v$ tends to infinity. In~the~vicinity of~the~singular $r=0$ line, there exist several additional hypersurfaces of~$r\Pv=0$ and~$r\Pu=0$. The~Cauchy horizon is present in~each spacetime within the~event horizon at $v=\infty$. For~the~nonvanishing $\beta$ parameter, each of~the~spacetimes forming during the~collapse with $\omega\geqslant -1$ consists of~a~spacelike singularity along $r=0$ surrounded by a~single apparent horizon $r\Pv=0$, whose location for large values of~advanced time indicates the~event horizon position in~the~spacetime. Neither additional apparent horizons nor the~Cauchy horizon exist in~the~spacetimes. In~the~ghost limit of~$\beta\neq 0$, i.e.,~for $\omega=-1.6$, the~future infinity is situated at large $u$ and~surrounded by an~apparent horizon $r\Pu=0$. Additionally, when $\beta=0.5$, from the~point of~coincidence of~the~two a~spacelike central singularity extends along $r=0$ towards smaller values of~$v$ and~larger retarded times. The~singularity is fully surrounded by an~$r\Pv=0$ apparent horizon.

The~dynamics of~both Brans-Dicke and~electrically charged scalar fields becomes more significant when the~singularity is approached, apart from the~case of~$\beta=0$ and~$\omega=-1.4$. The~variations of~the~field functions values are also considerable in~the~dynamical area of~the~spacetime, which is more apparent in~the~case of~a~complex scalar field. The~$u$-component of~the~electromagnetic field four-potential is most dynamical at large $v$ nearby the~event horizon, while the~charge function dynamics is analogous to the~complex scalar field behavior described above. In~all the~investigated cases, both the~Brans-Dicke and~the~electrically charged scalar fields can act as~time measurers in~dynamical spacetime regions of~high curvature, as~their constancy hypersurfaces are spacelike and~their values changes are monotonic there. The~existence of~separated points, at which single nonspacelike hypersurfaces of~constant field functions can reach the~singularity does not spoil this possibility. The~feasibility of~treating the~dynamical quantities as~`clocks' in~the~asymptotic spacetime regions as~the~values of~$v$ increase mainly depends on~the~value of~$\beta$. When the~parameter vanishes, the~Brans-Dicke field, as~well as~the~real and~imaginary parts of~the~complex scalar field function, along with its modulus and~the~charge function, are either spacelike or timelike near the~singularity for all values of~$\omega$. Hence, all these quantities are excluded from being time quantifiers. Moreover, $A_u$ is timelike along the~whole singularity, so it also cannot be used to measure time there. On~the~opposite, when $\beta\neq 0$ the~constancy hypersurfaces of~$\Phi$, $\textrm{Re}\;\phi$, $\textrm{Im}\;\phi$ and~$|\phi|$ are spacelike in~the~neighborhood of~the~singular $r=0$ and~they change monotonically there, except the~case of~$\beta=1$ and~$\omega=10$. This means that all these quantities can be used as~`clocks' in~the~vicinity of~the~singularity as~$v\to\infty$.

The~conducted analyses revealed that there does not exist a~good dynamical time measure in~the~spacetime region neighboring the~Cauchy horizon, because all the~examined quantities may be timelike in~this area. The~conclusions regarding the~Cauchy horizon, which is present in~the~spacetimes formed during the~Einstein collapse and~the~Brans-Dicke collapse with $\beta=0$, are similar to the~ones drawn in~the~case of~a~neutral scalar field collapse in~the~ghost regime with the~vanishing $\beta$. In~this case presented in~paper~I, measuring time with the~use of~dynamical quantities was also excluded nearby the~emerging Cauchy horizon.

The~dynamical quantities which can be used to~quantify time in~evolving gravitational systems are gathered in~table~\ref{tab:summary}, which summarizes the~investigations of~potential time measurements nearby the~singularities emerging during the~collapse within the~Einstein and~Brans-Dicke theories. The~outcomes of~the~research concerning the~charged collapse in~the~Brans-Dicke theory support the~conclusions about using dynamical quantities as~time variables within the~evolving spacetimes during dynamical gravitational evolutions of~coupled matter-geometry systems, which were formulated on~the~basis of~investigating the~neutral collapse in~paper~I. First, the~spacelike character of~the~constancy hypersurfaces of~the~quantities and~the~monotonicity of~their parametrization are not retained within the~whole spacetime. Second, there does not exist a~good time measure for the~areas nearby the~Cauchy horizons. Third, in~attempts to use dynamical quantities as~`clocks', special attention should be paid to the~values of~the~free evolution parameters, as~they can strongly influence this possibility.

\begin{table}[tbp]
\centering
\begin{tabular}{c|c|c|c|c|}
\cline{2-5}
&\multicolumn{4}{c|}{\multirow{2}{*}{scalar field collapse in the Einstein theory}}\\
&\multicolumn{4}{|c|}{}\\
\cline{2-5}
& \multicolumn{2}{c|}{neutral~\cite{NakoniecznaLewandowski2015-064031}} & \multicolumn{2}{c|}{charged} \\
\cline{2-5}
& \hspace{0.75cm}$\mathcal{D}$\hspace{0.75cm} & \hspace{0.75cm}$\mathcal{A}$\hspace{0.75cm} & \hspace{0.75cm}$\mathcal{D}$\hspace{0.75cm} & \hspace{0.75cm}$\mathcal{A}$\hspace{0.75cm} \\
\cline{2-5}
& \multicolumn{2}{c|}{$\phi$} & $\phi$, $q$ & $-$ \\
\cline{2-5}
&\multicolumn{4}{c|}{\multirow{2}{*}{scalar field collapse in the Brans-Dicke theory}}\\
&\multicolumn{4}{|c|}{}\\
\cline{2-5}
& \multicolumn{2}{c|}{neutral~\cite{NakoniecznaYeom2016-049}} & \multicolumn{2}{c|}{charged} \\
\cline{2-5}
& $\mathcal{D}$ & $\mathcal{A}$ & $\mathcal{D}$ & $\mathcal{A}$ \\
\hline
\multicolumn{1}{|c|}{$\beta=0$} & \multicolumn{2}{c|}{$\Phi$, $\phi$ $^{(1)}$} & $\Phi$, $\phi$, $q$ & $-$ \\
\hline
\multicolumn{1}{|c|}{$\beta=0.5$} & \multicolumn{2}{c|}{$\Phi$, $\phi$} & \multicolumn{2}{c|}{$\Phi$, $\phi$, $q$} \\
\hline
\multicolumn{1}{|c|}{$\beta=1$} & \multicolumn{2}{c|}{$\phi$} & \multicolumn{2}{c|}{$\Phi$, $\phi$, $q$ $^{(2)}$} \\
\hline
\end{tabular}
\caption{\label{tab:summary} Feasibility of~using dynamical quantities of~a~coupled matter-geometry system to~quantify time nearby the~singularity emerging during the~gravitational evolution. The~dynamical ($v\leqslant 20$) and~asymptotic ($v\to\infty$) spacetime regions are denoted as~$\mathcal{D}$ and~$\mathcal{A}$, respectively. The~scalar and~Brans-Dicke fields are $\phi$ and~$\Phi$, respectively, while $q$ denotes the~charge function~\eqref{eqn:q}. Notes:~$^{(1)}$except the~vicinity of~the~Cauchy horizon for~$\omega=-1.6$, $^{(2)}$except the~asymptotic region for~$\omega=10$.}
\end{table}

In~general, in~comparison to the~collapse of~a~neutral scalar field, the~presence of~the~electric charge in~the~spacetime modifies the~feasibility of~time quantification using the~evolving quantities in~the~following way. In~the~case of~uncoupled Brans-Dicke and~complex scalar fields, i.e.,~for $\beta=0$, the~charge spoils the~possibility of~measuring time with their use, while when the~fields are coupled, that is $\beta$ is not equal to zero, the~charge enhances~it. In~the~studied charged collapse, there also exist two more potential time measures, i.e.,~the~charge function and~the~nonzero component of~the~Maxwell field four-potential. In~the~context of~time quantification, the~former behaves analogously to the~complex scalar field dynamical characteristics. On~the~other hand, the~latter does not provide a~good time measure, as~its constancy hypersurfaces are timelike in~the~regions of~high curvature.

During the~gravitational collapse of~an~electrically charged scalar field, the~mass inflation phenomenon can appear in~the~spacetime. If so, a~super-Planckian surface develops outside the~true singularity~\cite{HwangYeom2011-064020}. Within the~region encompassed by it the~spacetime curvature reaches values excluding the~usage of~a~classical theory of~gravity. Hence, quantized gravity should be applied not around the~singularity in~this case, but around the~mass inflation super-Planckian surface. It~is possible that in~the~vicinity of~this hypersurface one of~the~quantities discussed above or their combination could provide a~good time measure. However, since the~proposed construction depends on~the~cutoff of~the~super-Planckian region, it requires more detailed studies on~the~region itself at first, as~the~determination of~its boundaries could be only arbitrary without any specific analyses. For~this reason, we~leave the~announced topic for future researches.

\acknowledgments

A.N. was partially supported by~the~Polish National Science Center grant no.~DEC-2014/15/ B/ST2/00089. D.Y. is supported by~Leung Center for~Cosmology and~Particle Astrophysics (LeCosPA) of~National Taiwan University (103R4000).


\appendix
\section{Numerical computations}
\label{sec:appendix}

The~solution of~the~evolution equations~\eqref{eqn:matrix2}, \eqref{eqn:sf1-dn}--\eqref{eqn:q-dn} was provided numerically with the~use of~an~enhanced version of~the~code prepared for the~needs of~paper~I. The~modules governing the~evolution of~the~gravitational, Brans-Dicke and~scalar fields were modified in~order to account for the~presence of~an~electrically charged scalar field instead of~a~neutral one. The~program was also supplied with a~module governing the~evolution of~the~Maxwell field. The~quantities $A_u$ and~$q$ were set as~equal to zero along the~initial $v=0$ hypersurface, because due to the~form of~the~evolving field, the~center of~the~shell was not affected by matter. The~$u$-component of~the~electromagnetic field four-potential and~the~charge function along the~initial $u=0$ hypersurface were calculated according to~\eqref{eqn:Au-dn} and~\eqref{eqn:q-dn}.

\begin{figure}[tbp]
\centering
\includegraphics[width=0.8\textwidth]{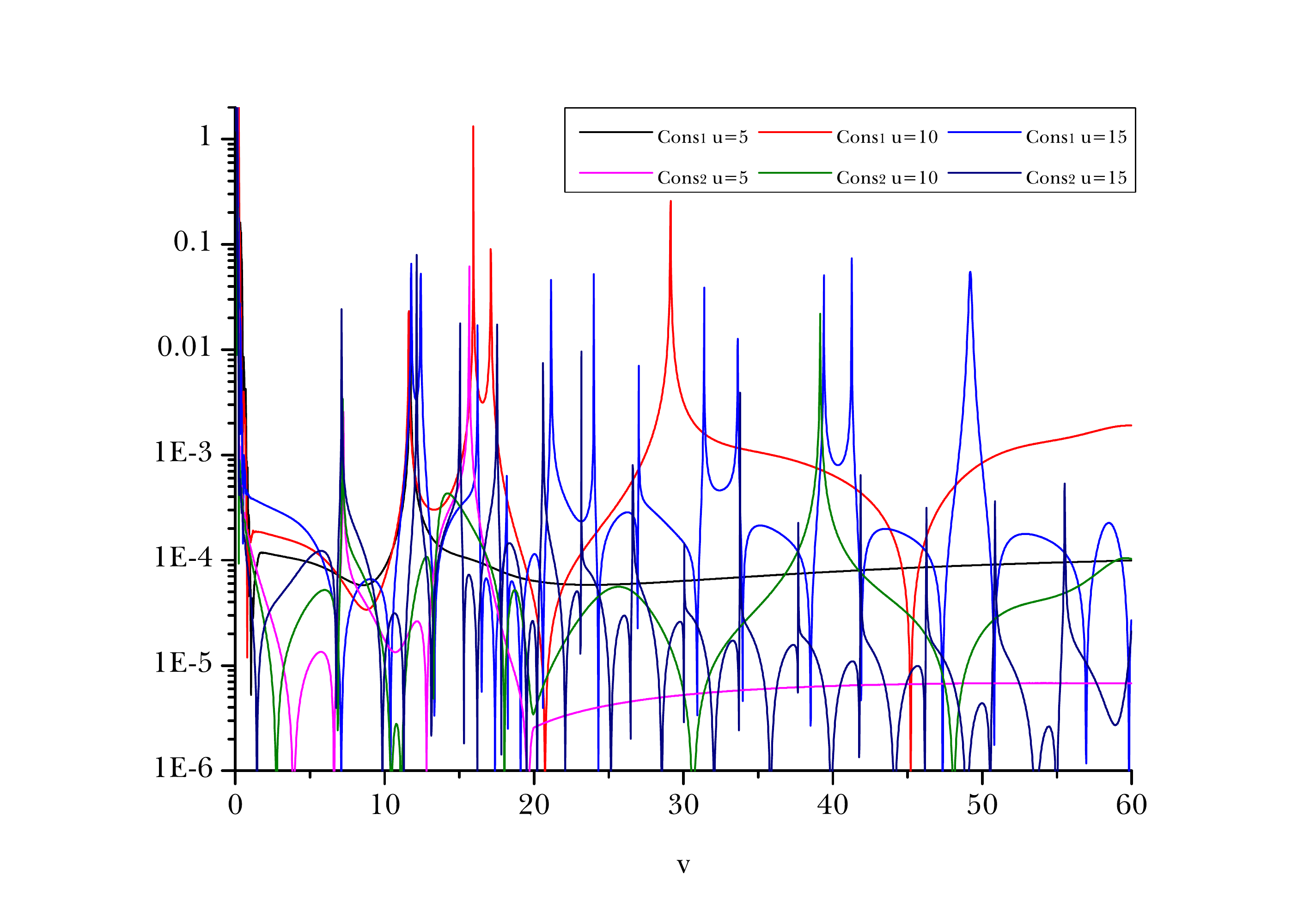}
\vspace{-0.35cm}
\caption{(color online) Monitoring of~the~constraints. The~values of~the~equations~\eqref{eqn:constr1} and~\eqref{eqn:constr2} were calculated along three null hypersurfaces of~constant~$u$ equal to~$5$, $10$ and~$15$.}
\label{fig:const}
\end{figure}

The~accuracy checks of~the~numerical code will be presented for a~sample evolution running for the~parameters $\beta=0$, $\omega=-1.4$ and~$e=0.3$. The~consistency of~the~computations was monitored during all evolutions using the~constraints~\eqref{constr1} and~\eqref{constr2}. Figure~\ref{fig:const} presents the~descendant quantities 
\ben\label{eqn:constr1}
\textrm{Cons}_1 &\equiv& \frac{2 \Big|r_{,uu} - 2fh + \frac{r}{2\Phi} \left( W\Pu - 2hW \right) + \frac{r\omega}{2\Phi^2}W^2
+ \frac{4\pi r}{\Phi}\widetilde{T}_{uu}^{EM}\Big|}{|r_{,uu}| + \Big|2fh - \frac{r}{2\Phi} \left( W\Pu - 2hW \right) - \frac{r\omega}{2\Phi^2}W^2 - \frac{4\pi r}{\Phi}\widetilde{T}_{uu}^{EM}\Big|}, \\
\label{eqn:constr2}
\textrm{Cons}_2 &\equiv& \frac{ \Big|r_{,vv} - 2dg + \frac{r}{2\Phi} \left( Z\Pv - 2dZ \right) + \frac{r\omega}{2\Phi^2}Z^2 + \frac{4\pi r}{\Phi}\widetilde{T}_{vv}^{EM}\Big|}{|r_{,vv}| + \Big|2dg - \frac{r}{2\Phi} \left( Z\Pv - 2dZ \right) - \frac{r\omega}{2\Phi^2}Z^2 - \frac{4\pi r}{\Phi}\widetilde{T}_{vv}^{EM}\Big|},
\een
calculated along three arbitrary null hypersurfaces for the~evolution with parameters specified above. The~values of~$\textrm{Cons}_1$ and~$\textrm{Cons}_2$ ought to be smaller than~$2$ in~order to satisfy the~constraints well (except the~case when $r_{,uu}$ or $r_{,vv}$ vanishes). As~can be inferred from the~plot, the~error is less than $1\%$ within almost the~whole integration domain. Since the~constraint equations are stable, the~simulations are consistent.

\begin{figure}[tbp]
\centering
\includegraphics[width=0.8\textwidth]{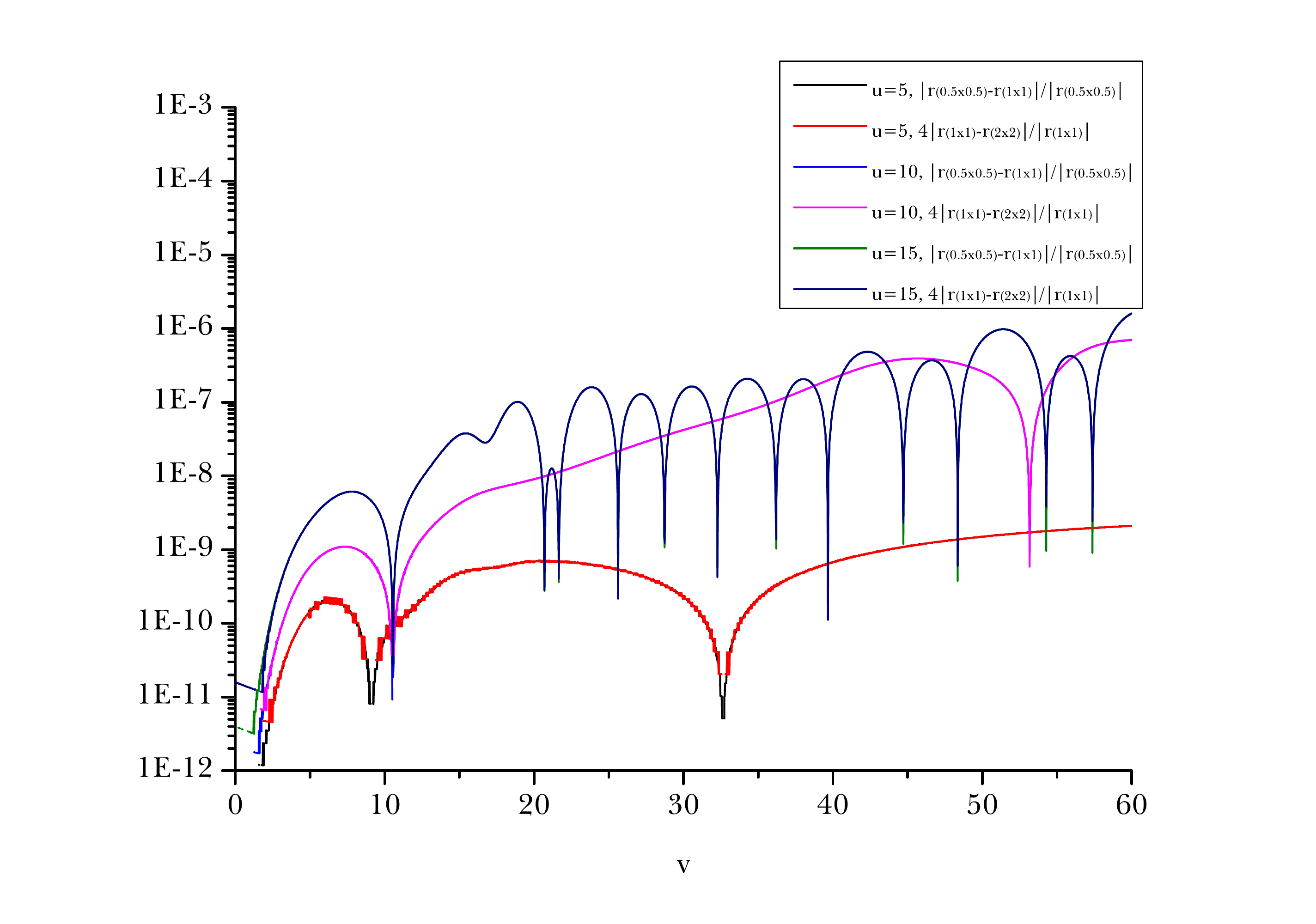}
\vspace{-0.35cm}
\caption{(color online) The~convergence of~the~code presented through the~prism of~the~values of~the~quantity $|r_{(k\times k)}-r_{(2k\times 2k)}|/|r_{(k\times k)}|$ with $k=0.5,1$ calculated at~the~same hypersurfaces of~constant~$u$ as~in~figure~\ref{fig:const}. $(k\times k)$ denotes the~resolution of~the~numerical grid, on~which the~computations were conducted.}
\label{fig:conv}
\end{figure}

The~outcome of~the~convergence check of~the~numerical code for the~sample evolution is presented in~figure~\ref{fig:conv}. The~values of~a~quantity constructed from the~$r$ function obtained on~two grids with a~quotient of~integration steps equal to~$2$ were calculated at three arbitrary $u=const.$ hypersurfaces. An~overlap between two profiles of~the~defined quantity at~each $u=const.$ was obtained when the~result from finer grids was multiplied by $4$. Thus, the~code displays a~second order convergence. The~discrepancy between each two profiles at~each constant~$u$ hypersurface is less than $10^{-4}\%$ except a~close vicinity of~the~singularity.

\bibliographystyle{JHEP}
\bibliography{measuringtime.bib}

\end{document}